\documentclass[aps,prb,preprint,groupedaddress,floatfix,showpacs]{revtex4-1}

\usepackage{epsfig}
\usepackage{amssymb}
\usepackage{amsmath}
\usepackage{bm}
\usepackage{hyperref}
\usepackage{enumerate}

\begin{document}
% Full title of the paper (Capitalized)
\title{Interplay between Point-Group Symmetries and the Choice of the Bloch Basis in Multiband Models}
%
% Authors (Add full first names)
\author{Stefan A. Maier,$^{\ast1}$ Carsten Honerkamp,$^1$ and Qiang-Hua Wang$^2$}
\affiliation{$^1$Institute for Theoretical Solid State Physics, RWTH Aachen University, D-52056 Aachen, Germany
\\ and JARA - FIT Fundamentals of Future Information Technology\\
$^2$ National Laboratory of Solid State Microstructures, Nanjing University, Nanjing 210093, China}
\date{Oct 29, 2013}

% Abstract
\begin{abstract}
We analyze the point-group symmetries of generic multiband tight-binding models with respect to the transformation properties of the effective interactions. While the vertex functions in the orbital language may transform non-trivially under point-group operations, their point-group behavior in the band language can be simplified by choosing a suitable Bloch basis.
 We first give two analytically accessible examples. Then we show that, for a large class of models, a natural Bloch basis exists, in which the vertex functions in the band language transform trivially under all point-group operations. As a consequence, the point-group symmetries can be used to reduce the computational effort in perturbative many-particle approaches such as the functional renormalization group.
\end{abstract}

% The fields PACS and MSC may be left empty or commented out if not applicable
\maketitle
\section{Introduction}

 Symmetries play an important role in solid state physics, as they may prohibit or protect certain
 features of a given system and often  help on the way to a simpler understanding.  \cite{tinkham-groupth}
 In the absence of spontaneous symmetry breaking, effective models can be formulated by analyzing which contributions
to the Hamiltonian are allowed for a given lattice geometry. In bosonic field theories, this often means that the leading low-energy behavior can be described by only a handful of coupling constants, or even less.
%In this way, both bosonic as well as fermionic degrees of freedom may be described. Famous examples of the latter are the Kane-Mele model \cite{kane-mele-model} for graphene and the single-band Hubbard model on a square lattice which has served as some kind of standard model for high $T_\mathrm{c}$ (see e.g.\ Ref.~\onlinecite{RVB}). In this spirit, also an $S_4$ symmetric model for the pnictide superconductors has been proposed recently.  \cite{S4-pnictides} The example of the Kane-Mele model, however, illustrates that models postulated on symmetry grounds may not always fully apply to the system for which they originally have been conceived. In graphene, a spin-orbit term is not prohibited by symmetries, but has been found to be negligibly small  in \textsl{ab initio} studies.  \cite{no-so-huertas,no-so-min} Nevertheless, the inclusion of the spin-orbit term has triggered a lot of interesting research (see, for example, Ref.~\onlinecite{kane-review}). So in order to decide which of the terms allowed by symmetries should be kept for a given system, one also has to resort to \emph{ab initio} techniques and/or to compare the predictions obtained from effective models in question to empirical data.

 For many-fermion models at low energy scales, the interactions for quasiparticle excitations near the Fermi level can however become rather rich in their momentum structure. This can e.g.\ be seen in perturbation theory for a single-band system like the basic Hubbard model. Here, particle-hole and particle-particle one-loop diagrams cause a substantial wavevector dependence of the effective interaction vertex, which encodes interesting properties like the spin-correlation length and the symmetry of the induced Cooper pairing correlations. These phenomena can be studied in low-order perturbation theory of partial summations, but possibly more satisfactorily by using renormalization group (fRG) techniques.  \cite{btw-review,metzner-rmp}
In most of these approaches, one works with an interaction function $V({\bf k}_1,{\bf k}_2,{\bf k}_3)$ that depends on three wavevectors in the Brillouin zone, with the fourth being fixed  by momentum conservation on the lattice. The discretization or expansion of the wavevector dependence of this interaction results in a large number of running couplings in the RG flow equations.
Hence, an integration of these flow equations can be numerically very demanding. A reduction of the computational effort by implementing the space-group symmetries appears therefore desirable. While translational symmetries are usually exploited by working in reciprocal space, the point-group symmetries shall be in the focus of this work.
In a single-band model, the point-group symmetry properties of the interaction function are straightforward, but in multiorbital problems (including problems with more than one site per unit cell), the orbital or band degree of freedom complicates the transformation properties of the terms in the (effective) action or Hamiltonian, such that a closer look is helpful. This is provided in the present article.

At present, there has been a series of fRG studies of multiband models working in the band picture for graphene systems, \cite{Honerkamp-honeycomb,BLG-frg,BLG-AF,TLG-frg,kiesel1} the pnictides, \cite{wang-2009,wang-2010,Thomale-Platt-pnictides,platt-first,pnictides-gap,sidplatt} or other two-dimensional systems.  \cite{kiesel2,sruo} In some of those works, the point-group symmetries have been exploited already, however without discussing the underlying formal structures.
 For a general multiband model for interacting fermions, however, this issue may require some care. Before we embark on
 this task, let us note in passing that there are other RG works that work in the orbital picture.  \cite{Vafek-BLG,ch-dec-graphene,ch-dec-kagome,sruo} For those studies, the transformation behavior in the band language discussed below is less
 relevant, but the symmetry properties in the orbital picture described in the first part will still apply.

To get started, let us state that in a second-quantization language, a many-particle Hamiltonian is expressed in terms of field operators which can be called \emph{auxiliary} quantities. However, the presence or absence of a physical symmetry manifests itself in \emph{observable} quantities. Although the energy of the system is such an observable,
the vertex functions of a many-particle model are auxiliary quantities in general, since they play the role of coefficients in an expansion in auxiliary quantities. Of course, one should in principle be able to find transformation rules according to which the physical symmetries manifest themselves in auxiliary quantities such as the vertex functions. This is of course a similar issue as the possible non-trivial transformation properties of specific wavefunctions in elementary quantum mechanics, e.g. in the case of rotational symmetry, while the observables should reflect the symmetry in a trivial way. Another textbook example that covers issues similar to those discussed below is the construction of the transformation law of the Dirac 4-spinor under Lorentz-transformation such that the Lagrangian remains invariant. Below we construct the corresponding transformation for the field operators in a multiorbital problem, such that the Hamiltonian (or the Lagrangian) density in the orbital picture and later in the band picture remains invariant under the point group.

 For symmetries other than those in the point-group, symmetry constraints on fermionic vertex functions have been derived from such transformation rules in the fRG literature, \cite{salm_hon_2001,husemann_2009,eberlein_param} particle-hole and time-reversal symmetries being examples. In these cases, the corresponding symmetries could be implemented due to
a simple form of these constraints.
At this point, however, it is not clear why also the point-group transformation rules of the vertex functions should take on a simple form which would allow for a reduction of the numerical effort.
 Since we are dealing with auxiliary quantities, even further complications may arise. If a theory is expressed
 in auxiliary quantities, there may be \emph{gauge or basis transformations} affecting the auxiliary, but not the observable quantities. In this work, such transformations will have the character of basis rather than of gauge transformations, since not only the fields, but also the vertex functions will be affected. The precise form of the point-group transformation rules for the vertex functions may consequently be basis-dependent in general.

 Despite the equivalence of all possible bases, one of them may be more convenient than another in a particular context. The choice of maximally localized Wannier functions, for example, may be very helpful.  \cite{wannier-rmp}
 For a second-quantized tight-binding Hamiltonian, the precise form of the vertex functions depends on the Wannier basis chosen. In particular, weakly localized Wannier orbitals will result in long-range hopping terms.
 In this work, we will exploit the freedom in the choice of a momentum-dependent phase of the Bloch states in the band language.

 One may therefore wonder, whether the phase of the Bloch state in the band language can be fixed in such a way that the point-group transformation rules for the vertex functions take on a simple form which allows for further progress and makes the symmetry explicit.
 We will show that, for a large class of tight-binding models, there always exists a \emph{natural Bloch basis} with
 transformation rules for the vertex functions that only affect the momentum quantum numbers.
 Furthermore, the existence of such natural bases also gives rise to a simple transformation rule of the vertex functions in most non-natural gauges.
 In real space, the choice of the phases of the Bloch states  corresponds to the above mentioned freedom in the localization properties of Wannier functions. Hence, the interpretation of a real-space formulation requires some care.

 In order to point this out, we start with a general fermionic many-particle Hamiltonian expressed in a second-quantization language in Sec.~\ref{sec:gen-ham}. In Sec.~\ref{sec:Wannier}, it is first expressed in a basis of Wannier states that hybridize in the one-particle part of the Hamiltonian. We then switch to a reciprocal space description
in Sec.~\ref{sec:Bloch}, where the basis states are Bloch states, which again hybridize.
 We also consider a non-hybridizing Bloch basis, i.e.\ a description in terms of bands instead of orbitals.

 Before analyzing the point-group behavior of the general Hamiltonian, we first study two examples where the transformation from orbitals to bands is analytically accessible.
 These are the Emery model without oxygen-oxygen hopping discussed in Sec.~\ref{sec:Emery} and an extended Hubbard model for fermions on a honeycomb lattice which will be analyzed in Sec.~\ref{sec:graphene}. The point groups of these models are $C_{4v} $ and $ C_{6v} $, respectively.
 For these two models, we proceed as follows: First we give the point-group transformation rules for the vertex functions in the orbital language in Secs.~\ref{sec:hidden-c4v} and \ref{sec:hidden-c6v}.
 We then switch to the band language in Secs.~\ref{sec:Emery-natural} and \ref{sec:graphene-natural} and observe that the vertex functions transform trivially under point-group operations in a particular non-hybridizing Bloch basis.

 In Sec.~\ref{sec:general}, we return to the general case in order to see whether a larger class of models enjoys
this property. In Sec.~\ref{sec:trans-rule}, we point out that not only the momentum, but also the orbital quantum numbers of the fields can be affected by a point-group operation for our general Hamiltonian in the orbital language. In Sec.~\ref{sec:gen-band}, we switch to the band language and show that the one-particle vertex function then has a trivial point-group behavior. Moreover, we find that, for the one-particle part of the Hamiltonian, switching to
another basis of non-hybridizing Bloch states corresponds to a mapping between equivalent representations of the point group. In the presence of interactions, this only holds if the new Bloch basis does not violate a natural-basis condition. In Sec.~\ref{sec:gen-natural}, we elaborate on the properties of such natural Bloch bases -- in particular on the trivial point-group behavior of the two-particle and higher vertex functions and on the consequences of their existence for calculations in non-natural bases.
Finally, we conclude by an outlook on possible applications of our findings in Sec.~\ref{sec:concl}.
\section{General tight-binding Hamiltonian} \label{sec:gen-ham}

\subsection{Wannier basis} \label{sec:Wannier}

Consider  a general many-particle Hamiltonian
 \begin{equation}
  H = H_0 + \sum_{n= 2}^m H^\mathrm{int}_n
 \end{equation} 
 for electrons in some solid in $D$ dimensions with a one-particle part $H_0$ and two- up to $m$-particle interaction terms $H^\mathrm{int}_n$.
 (Although three-particle and higher interactions are generically absent, i.e.\ although one usually has $m=2$, some statements that will be made in this work hold as well if these higher-order interaction terms are present.)
 Let us further suppose that, from some \textsl{ab initio} method, we have a basis set of (fairly well localized)
 Wannier functions at hand.
 If the above model Hamiltonian describes $l$ of these orbitals per unit cell, these states can be labeled in the
 following way. The one-particle state
\begin{equation}
 \left| \psi_a^\alpha (\mathbf{R}) \right\rangle
\end{equation} 
 is associated with the $\alpha$th orbital in the direct unit cell with center $\mathbf{R}$.
 This convention seems not to be very widespread in the literature, but has the advantage that the position quantum numbers $\mathbf{R} $ of Wannier orbitals corresponding
to electronic orbitals on different atoms all live on the same Bravais lattice.
 The subscript  $a$ denotes a collection of other quantum numbers, which usually include the spin projection.
 Note that the position $\mathbf{R}$ plays the role of a quantum number and should not be confused with the argument $\mathbf{r}$ of the wavefunction
 $ \left\langle \mathbf{r} \left| \psi_a^\alpha (\mathbf{R}) \right. \right\rangle $
 in position representation.
 In proceeding towards a second quantization language, Slater determinant  $n$-particle states
\begin{equation}
 \left| \psi_{a_1}^{\alpha_1} (\mathbf{R}_1) \dots \psi_{a_n}^{\alpha_n} (\mathbf{R}_n) \right\rangle
\end{equation} 
are written as excitation of the vacuum $ \left| 0 \right\rangle$ with
 $ \Psi_a^\alpha (\mathbf{R}) \left| 0 \right\rangle = 0 $
 according to
\begin{equation}
 \left| \psi_{a_1}^{\alpha_1} (\mathbf{R}_1) \dots \psi_{a_n}^{\alpha_n} (\mathbf{R}_n) \right\rangle
= {\Psi_{a_1}^{\alpha_1}}^\dagger (\mathbf{R}_1) \dots {\Psi_{a_n}^{\alpha_n}}^\dagger (\mathbf{R}_n) \left| 0 \right\rangle \, .
\end{equation} 
 The field operators $ \Psi_a^\alpha (\mathbf{R}) $ and $ {\Psi_a^\alpha}^\dagger (\mathbf{R}) $  obey
 the canonical commutation relations for fermions and consequently
\begin{equation}
 \Psi_{a_1}^{\alpha_1} (\mathbf{R}_1) \left| \psi_{a_1}^{\alpha_1} (\mathbf{R}_1) \, \psi_{a_2}^{\alpha_2} (\mathbf{R}_2) \dots \psi_{a_n}^{\alpha_n} (\mathbf{R}_n) \right\rangle
 = \left| \psi_{a_2}^{\alpha_2} (\mathbf{R}_2) \dots \psi_{a_n}^{\alpha_n} (\mathbf{R}_n) \right\rangle \, .
\end{equation} 
 In a second-quantized language, the Hamiltonian now reads as \cite{Fetter-Walecka}
\begin{align} \notag
 H &= \sum_{a_1,a_2} \sum_{\alpha_1,\alpha_2} \sum_{\mathbf{R}_1,\mathbf{R}_2}
\mathcal{T} \left( a_1,\alpha_1,\mathbf{R}_1;a_2,\alpha_2,\mathbf{R}_2\right) {\Psi_{a_1}^{\alpha_1}}^\dagger (\mathbf{R}_1) \,\Psi_{a_2}^{\alpha_2} (\mathbf{R}_2) \\ \notag
  &\quad + \sum_{n= 2}^m \sum_{a_1,\dots,a_{2n}} \sum_{\alpha_1,\dots,\alpha_{2n}} \sum_{\mathbf{R}_1,\dots,\mathbf{R}_{2n}}
\mathcal{U}_n \left( a_1,\alpha_1,\mathbf{R}_1;\dots;a_{2n},\alpha_{2n},\mathbf{R}_{2n}\right) \\ & \qquad
 \times {\Psi_{a_1}^{\alpha_1}}^\dagger (\mathbf{R}_1) \dots {\Psi_{a_n}^{\alpha_n}}^\dagger (\mathbf{R}_n) \, \Psi_{a_{n+1}}^{\alpha_{n+1}} (\mathbf{R}_{n+1}) \dots \Psi_{a_{2n}}^{\alpha_{2n}} (\mathbf{R}_{2n})
\end{align} 
with the vertex functions
\begin{align}
 \mathcal{T} \left( a_1,\alpha_1,\mathbf{R}_1;a_2,\alpha_2,\mathbf{R}_2\right) & = \left\langle \left. \psi_{a_1}^{\alpha_1} (\mathbf{R}_1) \right| H_0 \left|  \psi_{a_2}^{\alpha_2} (\mathbf{R}_2)  \right. \right\rangle \\
\mathcal{U}_n \left( a_1,\alpha_1,\mathbf{R}_1;\dots;a_{2n},\alpha_{2n},\mathbf{R}_{2n}\right)
&= \left\langle \left. \psi_{a_1}^{\alpha_1} (\mathbf{R}_1)\dots \psi_{a_n}^{\alpha_n} (\mathbf{R}_n) \right| H_n^\mathrm{int} \left| \psi_{a_{2n}}^{\alpha_{2n}} (\mathbf{R}_{2n}) \dots  \psi_{a_{n+1}}^{\alpha_{n+1}} (\mathbf{R}_{n+1})  \right. \right\rangle \, .
\end{align} 
Clearly, the precise form of these vertex functions depends on the choice of the basis. If a fairly localized Wannier basis has been chosen, long-ranged terms in the vertex functions are typically negligible.
(In the two-particle part of an effective Hamiltonian, for example, this corresponds to a screened Coulomb interaction between the electrons.)
 Note that, in general, the one-particle vertex function $\mathcal{T}$ contains hopping terms between different
species of orbitals.
\subsection{Hybridizing and non-hybridizing Bloch bases} \label{sec:Bloch}

 Let us now switch to reciprocal space, i.e.\ to the basis of Bloch states
\begin{equation}
 \left | \phi^\alpha_a (\mathbf{k}) \right\rangle = \sum_\mathbf{R} e^{i \mathbf{k} \cdot \mathbf{R}} \,
 \left| \psi_a^\alpha (\mathbf{R}) \right\rangle
= \left[ \sum_\mathbf{R} e^{i \mathbf{k} \cdot \mathbf{R}} \, {\Psi_a^\alpha}^\dagger (\mathbf{R}) \right] \left| 0 \right\rangle
= {\Psi_a^\alpha}^\dagger (\mathbf{k}) \left| 0 \right\rangle
\end{equation} 
with wavevectors $\mathbf{k}$ in the first Brillouin zone (BZ) $\mathbb{B}$ and new field operators
$\Psi_a^\alpha (\mathbf{k}) $. The sum over $\mathbf{R}$ in the Fourier transform defining these new fields
runs over the centers of the unit cells of the direct lattice.
In this new basis, the Hamiltonian reads as
\begin{equation} \label{eqn:gen-model}
 H= \sum_a \int_\mathbb{B} \! d\mathbf{k} \, \Psi_a^\dagger (\mathbf{k}) \, \mathcal{H}_0 (\mathbf{k}) \, \Psi_a (\mathbf{k}) + H_\mathrm{int} \left[ \Psi^\dagger, \Psi \right] \, .
\end{equation} 
The one-particle part is given in matrix notation with $l$-component pseudo-spinors $ \Psi_a (\mathbf{k}) $, called \emph{orbitors} in the following.
In Eq.~(\ref{eqn:gen-model}), we have assumed that the quadratic (one-particle) part $ \mathcal{H}_0 $ of the Hamiltonian does not depend on these additional quantum numbers. %This assumption seems realistic.
 For example, if $a$ denotes the spin-projection
quantum number, $\mathcal{H}_0$ is independent of this quantum number for a SU(2) symmetric theory.
 If the SU(2) invariance is broken, it may be advantageous to include components with different spin projection quantum numbers into the orbitor, i.e.\ in the quantum numbers $ \alpha $ instead of in $a$.

 The interaction with coupling functions $ \mathcal{V} $ now reads as
\begin{align} \notag
 H_\mathrm{int} & = \sum_{n=2}^m \int_\mathbb{B} \! d \mathbf{k}_1  \dots d \mathbf{k}_{2n} \,\,
  \mathcal{V}_{a_1, \dots, a_{2n}}^{\alpha_1,\dots,\alpha_{2n}}  ( \mathbf{k}_1, \dots, \mathbf{k}_n; \mathbf{k}_{n+1}, \dots, \mathbf{k}_{2n}) \\ & \quad  \times
 {\Psi_{a_1}^{\alpha_1}}^\dagger (\mathbf{k}_1) \dots {\Psi_{a_n}^{\alpha_n}}^\dagger (\mathbf{k}_n) \,\,
 {\Psi_{a_{n+1}}^{\alpha_{n+1}}}(\mathbf{k}_{n+1}) \dots {\Psi_{a_{2n}}^{\alpha_{2n}}} (\mathbf{k}_{2n}) \, ,
\end{align} 
where the summation over the orbital indices $ \alpha_i $ and other quantum numbers $ a_i $ is implicit.

 For the evaluation of observables or the calculation of Feynman diagrams as they, e.g., appear on the right-hand side of RG flow equations, it seems rewarding to work in a basis in which $ \mathcal{H}_0 (\mathbf{k}) $ is diagonal. This way, the effort invested
in index summations at internal legs of vertices can be reduced.%
\footnote{For other reasons, fRG calculations in the orbital language can still be viable and maybe even preferable in some cases, cf.\ for example Refs.~\onlinecite{ch-dec-kagome,ch-dec-graphene,sruo}.}
 This can be accomplished  by a unitary transformation $ u (\mathbf{k}) $ of the orbitor $ \Psi_a (\mathbf{k}) $ in the orbital picture to the band picture with pseudo-spinors
\begin{equation} \label{eqn:intro-u}
 \bm{\chi}_a (\mathbf{k}) = u( \mathbf{k}) \, \Psi_a (\mathbf{k}) \, ,
\end{equation} 
where $u(\mathbf{k}) $ is a $l\times l$ matrix with components $ u_{\alpha \beta} $ relating the $ \beta $th orbital to the $ \alpha$th band.
 The hybridizing one-particle Bloch basis states are consequently transformed to non-hybridizing ones
\begin{equation}
 \left| \chi_a^\alpha (\mathbf{k}) \right\rangle = \sum_\beta u_{\alpha,\beta} (\mathbf{k}) \, \left| \phi_a^\beta (\mathbf{k}) \right\rangle \, .
\end{equation} 
 The dispersion of the $\alpha$th band is then given by the component $ \mathcal{B}_{\alpha \alpha} (\mathbf{k}) $ of the diagonal matrix
\begin{equation}
 \mathcal{B} (\mathbf{k}) = u (\mathbf{k}) \, \mathcal{H}_0 (\mathbf{k}) \, u^\dagger (\mathbf{k})
\end{equation} 
in the quadratic part
\begin{equation}
 H_0 = \sum_a \int \! d\mathbf{k} \, \bm{\chi}^\dagger_a (\mathbf{k}) \, \mathcal{B} (\mathbf{k}) \, \bm{\chi}_a (\mathbf{k})
\end{equation} 
of the Hamiltonian.
%Typically even for metals, only some of the bands are \emph{conduction bands}, i.e.\ they a Fermi surface (FS) where $ \mathcal{B}_{\alpha \alpha} (\mathbf{k}) =0 $. The other bands, in contrast, are then separated from the Fermi surface by an energy gap $ \Delta E $, i.e.\ one has $ \left| \mathbf{B}_{\alpha \alpha} (\mathbf{k}) \right| \geq \Delta E $ on the whole BZ for these \emph{valence bands}.
%
Let us now rewrite also the interacting part of the Hamiltonian in the band language. In terms of the new fields $ \bm{\chi} $, it reads as
\begin{align} \notag
 H_\mathrm{int} & = \sum_{n=2}^m \int_\mathbb{B} \! d \mathbf{k}_1  \dots d \mathbf{k}_{2n} \,\,
  \mathcal{F}_{a_1, \dots, a_{2n}}^{\alpha_1,\dots,\alpha_{2n}}  ( \mathbf{k}_1, \dots, \mathbf{k}_n; \mathbf{k}_{n+1}, \dots, \mathbf{k}_{2n}) \\
 & \quad  \times
 {\bm{\chi}_{a_1}^{\alpha_1}}^\dagger (\mathbf{k}_1) \dots {\bm{\chi}_{a_n}^{\alpha_n}}^\dagger (\mathbf{k}_n) \,\, {\bm{\chi}_{a_{n+1}}^{\alpha_{n+1}}}(\mathbf{k}_{n+1}) \dots {\bm{\chi}_{a_{2n}}^{\alpha_{2n}}} (\mathbf{k}_{2n})
\end{align} 
with the $n$-particle coupling function
\begin{align} \notag
  \mathcal{F}_{a_1, \dots, a_{2n}}^{\alpha_1,\dots,\alpha_{2n}}   &( \mathbf{k}_1, \dots, \mathbf{k}_n; \mathbf{k}_{n+1}, \dots, \mathbf{k}_{2n}) =
  \mathcal{V}_{a_1, \dots, a_{2n}}^{\beta_1,\dots,\beta_{2n}}  ( \mathbf{k}_1, \dots, \mathbf{k}_n; \mathbf{k}_{n+1}, \dots, \mathbf{k}_{2n}) \\ \label{eqn:band-inter}
 & \quad  \times u_{\alpha_1 \beta_1} (\mathbf{k}_1) \dots u_{\alpha_n \beta_n} (\mathbf{k}_n) \, u_{\alpha_{n+1} \beta_{n+1}} (\mathbf{k}_{n+1})^\ast  \dots u_{\alpha_{2n} \beta_{2n}} (\mathbf{k}_{2n})^\ast \, .
\end{align} 
 One can observe that the momentum dependence of the $n$-particle coupling function is modulated by the
 (wavevector-dependent) transformation matrix elements $ u_{\alpha \beta} (\mathbf{k})$. In particular,
if the interaction is completely wavevector-independent in the orbital language, a nontrivial momentum dependence
 emerges in the band picture. An on-site Hubbard term, for example, is then rendered nonlocal by this so-called
 \emph{orbital makeup}, \cite{tmaier} which may have a considerable impact on the phase diagram
of multiband models for unconventional superconductors, for example.
 It is also believed to account for the differences between the phase diagrams of extended Hubbard models on the honeycomb and the kagome lattices.  \cite{kiesel2,ch-dec-kagome}
 In addition, it lends a non-trivial behavior under point-group operations to the interaction. In this work, we discuss how this behavior can be simplified.

In this place, one might also want to perform an inverse Fourier transform on the basis states
$ \left| \chi^\alpha_a \right\rangle $, at least for some bands $\alpha$ in the low-energy sector. This would lead to \emph{non-hybridizing} Wannier states. The localization of these new states should, however, be expected to be worse than for the hybridizing Wannier states $ \left| \Psi \right\rangle $. This, e.g., also happens when 8-band models for iron arsenides in the basis of both Fe$3d$ and As$4p$ Wannier states are reduced to 5-band models with $d$-like Wannier orbitals on the Fe sites. These low-energy effective
 orbitals typically extend somewhat more on the As sites, and hence are less strongly localized than the previous Fe orbitals.

We will henceforth use the following nomenclature. Instead of Wannier or Bloch states, we will also speak of
 real and reciprocal space descriptions, respectively. The expansion in states which do not hybridize at the one-particle level will be referred to as the \emph{band language}, while in the \emph{orbital language} these states hybridize. The basis transformations considered in this work correspond to the multiplication of the Bloch states or field operators in reciprocal space by a phase, i.e.\
\begin{equation}
 \Psi_a^\alpha (\mathbf{k})  \to e^{i \vartheta_\alpha (\mathbf{k}) } \, \Psi_a^\alpha (\mathbf{k})
\end{equation} 
 in the orbital language and
\begin{equation}
 \bm{\chi}_a^\alpha (\mathbf{k})  \to e^{i \varphi_\alpha (\mathbf{k}) } \, \bm{\chi}_a^\alpha (\mathbf{k})
\end{equation} 
 in the band language. The latter freedom is sometimes referred to as a ${\bf k}$-local U(1) invariance of the electronic structure. Since these transformations both correspond to convolution operations in real space,
 they may significantly affect the localization properties of both hybridizing and non-hybridizing Wannier states.
 Therefore, an interpretation in real space has to be made with care, keeping these phases in mind.

Throughout this work, we will only consider phase transformations that are independent of the additional quantum
numbers $a$.
One might be tempted to call these transformations a Bloch regauging. We will avoid using this term, since
 the vertex functions in a second-quantized language transform non-trivially.
 Therefore, there is no redundancy of the state description
 and the above transformations are strictly speaking not gauge transformations.
\section{Fourfold symmetry: Emery model without oxygen-oxygen hopping} \label{sec:Emery}

 \subsection{Model Hamiltonian}

\begin{figure}
 \centering
 \includegraphics[width=0.5\linewidth]{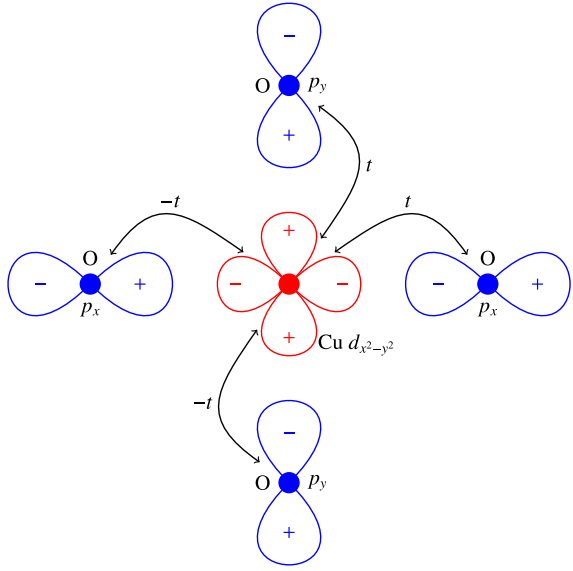}
 \caption{Orbitals of the Emery model depicted for one unit cell of the direct lattice (cf.\ Ref.~\onlinecite{fulde}, Fig.~14.12). Note that the signs of the electronic orbitals violate  the point-group ($C_{4v}$) symmetry of the underlying
lattice. The $+$ and $-$ signs in this Figure correspond to the sign of the orbital wave functions at the respective positions. The black arrows correspond to
 the hopping terms in the Hamiltonian in Eq.~(\ref{eqn:emery-realsp}).} \label{fig:CuO-plane}
\end{figure}

As a first example, let us consider the Emery model devised for the description of the Copper-oxide planes of the high-$T_\mathrm{c}$
cuprates.  \cite{Emery}
This model includes the Cu $3d$-orbitals at the center of the Wigner-Seitz cell as well as the oxygen $2p$-orbitals at the boundaries of the unit cell with fields $ d $ and $ p_x$ or $ p_y$, respectively (cf.\ Fig.~\ref{fig:CuO-plane}).
 In order to keep our calculations simple, we restrict ourselves to the (probably unrealistic \cite{andersen-YBCO,weber-2008,weber-2010a,weber-2010b}) case of vanishing oxygen-oxygen
hopping.

 For simplicity, let us first consider the quadratic part $ H_0 $ of the Emery Hamiltonian $ H_0 + H_\mathrm{int} $ and start from a real space formulation, i.e.\ form a basis of hybridizing Wannier states. For the labeling of the atoms, we chose the following
convention. While a particular Copper atom is located at the center $\mathbf{R}$ of some direct unit cell, the neighboring oxygen atoms in positive $x$ and $y$ direction also belong to the unit cell with center $\mathbf{R}$. The nearest oxygen atoms in negative $x$ and $y$ direction, in contrast, belong to neighboring cells. The one-particle Hamiltonian then reads as
\begin{align} \notag
 H_0 & =  \epsilon \sum_{\mathbf{R},\sigma,\nu} p^\dagger_{\nu,\sigma} (\mathbf{R}) \, p_{\nu,\sigma} (\mathbf{R})
+ t \sum_{\mathbf{R},\sigma,\nu} \left[ d^\dagger_\sigma (\mathbf{R}) \, p_{\nu,\sigma} (\mathbf{R}) + p^\dagger_{\nu,\sigma} (\mathbf{R}) \, d_\sigma (\mathbf{R}) \right]  \\ \label{eqn:emery-realsp}
& \quad - t \sum_{\mathbf{R},\sigma,\nu} \left[ d^\dagger_\sigma (\mathbf{R}) \, p_{\nu,\sigma} (\mathbf{R}- \hat{\bm{\nu}}) +
p^\dagger_{\nu,\sigma} (\mathbf{R}- \hat{\bm{\nu}}) \, d_\sigma (\mathbf{R})  \right]  \, ,
\end{align} 
where $ \hat{\bm{\nu}} $ represents the primitive lattice vector pointing in positive $ \nu $-direction,
i.e. either in $x$- or $y$-direction. The length of these primitive lattice vectors is just the distance between
 neighboring Copper atoms, which we henceforth set to unity.
 In Eq.~(\ref{eqn:emery-realsp}), $ \epsilon$ denotes the energy separation of the copper $d$- and oxygen $p$-orbitals and $t$ corresponds to the absolute value
 of the transfer integrals between neighboring copper and oxygen atoms.
 If one now switches to a Bloch representation
 with wavevectors $\mathbf{k} $ in the first Brillouin zone $ \mathbb{T} = [ -\pi,\pi) \times [ - \pi,\pi) $
 according to
\begin{align}
 d_\sigma (\mathbf{R}) & = \int_\mathbb{T} \! d\mathbf{k} \, e^{i \mathbf{k} \cdot \mathbf{R}} \, \tilde{d}_\sigma (\mathbf{k}) \\
 p_{\nu,\sigma} (\mathbf{R}) & = \int_\mathbb{T} \! d\mathbf{k} \, e^{i \mathbf{k} \cdot \mathbf{R}} \, \tilde{p}_{\nu,\sigma} (\mathbf{k})  \, ,
\end{align} 
we obtain
\begin{align} \notag
 H_0 & =  \epsilon \sum_{\sigma,\nu} \int_\mathbb{T}  \! d\mathbf{k} \,\,  \tilde{p}^\dagger_{\nu,\sigma} (\mathbf{k}) \, \tilde{p}_{\nu,\sigma} (\mathbf{k})
+ t \sum_{\sigma,\nu} \int_\mathbb{T}  \! d\mathbf{k}  \left[ \tilde{d}^\dagger_\sigma (\mathbf{k}) \, \tilde{p}_{\nu,\sigma} (\mathbf{k}) + \tilde{p}^\dagger_{\nu,\sigma} (\mathbf{k}) \, \tilde{d}_\sigma (\mathbf{k})  \right]  \\ \label{eqn:proper}
& \quad - t \sum_{\sigma,\nu} \int_\mathbb{T}  \! d\mathbf{k} \left[ e^{-i \mathbf{k} \cdot \hat{\boldsymbol \nu}} \, \tilde{d}^\dagger_\sigma (\mathbf{k}) \, \tilde{p}_{\nu,\sigma} (\mathbf{k}) +
e^{+i \mathbf{k} \cdot \hat{\boldsymbol \nu}} \, \tilde{p}^\dagger_{\nu,\sigma} (\mathbf{k}) \, \tilde{d}_\sigma (\mathbf{k}) \right] \, .
\end{align} 
 This expression is now cast into the form
\begin{equation}
  H_0  = \sum_{\sigma} \int_\mathbb{T} \! d\mathbf{k} \, \tilde{\Psi}_\sigma^\dagger (\mathbf{k}) \, \tilde{\mathcal{H}}_0 (\mathbf{k}) \, \tilde{\Psi}_\sigma (\mathbf{k})
\end{equation} 
 with orbitors
\begin{equation}
\tilde{\Psi}_\sigma (\mathbf{k}) = \left( \begin{array}{c} \tilde{d}_\sigma (\mathbf{k}) \\ \tilde{p}_{x,\sigma} (\mathbf{k}) \\ \tilde{p}_{y,\sigma} (\mathbf{k}) \end{array}  \right) \, .
\end{equation} 
 The one-particle coupling function $ \tilde{\mathcal{H}}_0 (\mathbf{k}) $ then clearly is $2 \pi$ periodic in both directions and hence no discontinuities occur at the boundary of the BZ. Moreover,  $ \tilde{\mathcal{H}}_0 (\mathbf{k}) $ has complex entries and hence is hermitian, but not symmetric.

 For numerical calculations, it may, however, be convenient to have only real valued coupling functions. This can be accomplished by a regauging of the fields: If Eq.~(\ref{eqn:proper}) is expressed in terms of new orbitors
\begin{equation}
\Psi_\sigma (\mathbf{k}) = \left( \begin{array}{c} \tilde{d}_\sigma (\mathbf{k}) \\ e^{-i \left( \mathbf{k} \cdot \hat{\mathbf{x}} - \pi \right) /2} \, \tilde{p}_{x,\sigma} (\mathbf{k}) \\ e^{-i \left( \mathbf{k} \cdot \hat{\mathbf{y}} - \pi \right) /2} \, \tilde{p}_{y,\sigma} (\mathbf{k}) \end{array}  \right) \, ,
\end{equation} 
 we obtain the one-particle coupling function $\mathcal{H}_0 (\mathbf{k}) $ given by
\begin{align} \label{eqn:1pHam}
 \mathcal{H}_0 (\mathbf{k}) & = \left( \begin{array}{ccc} 0 & 2 t \, \sin (k_x/2) &  2 t \, \sin (k_y/2) \\
 2 t \, \sin (k_x/2)& \epsilon& 0 \\
 2 t \, \sin (k_y/2) & 0 & \epsilon \end{array} \right)  \, ,
\end{align} 
which is real valued.
This comes at the price of loosing the continuity of the one-particle coupling function at the boundary of the BZ. We will henceforth call a basis of hybridizing Bloch states with a continuous one-particle coupling function a \emph{proper} one. Note that the improper basis of Eq.~(\ref{eqn:1pHam}) is related to the proper one by a unitary transformation that is discontinuous in $\mathbf{k}$. Since the Hamiltonian is local in momentum
space, this basis transformation corresponds to a unitary transformation of the one-particle coupling function $\mathcal{H}_0 (\mathbf{k}) $.

In real space, the interacting part of the Hamiltonian reads
\begin{align} \notag
 H_\mathrm{int}  & =  U_d \sum_\mathbf{R} : n_{d,\uparrow} (\mathbf{R}) \, n_{d,\downarrow} (\mathbf{R}) : + U_p \sum_{\mathbf{R},\nu} : n_{p,\nu,\uparrow} (\mathbf{R}) \, n_{p,\nu,\downarrow} (\mathbf{R}) : \\ \label{eqn:Hint}
  & \quad  + U_{pd} \sum_{\mathbf{R},\nu} : n_d (\mathbf{R}) \, n_{p,\nu} (\mathbf{R}) +n_d (\mathbf{R}) \, n_{p,\nu} (\mathbf{R}-\hat{\bm{\nu}}) : \, .
\end{align} 
In this equation, $:O:$ denotes the normal ordering of an operator product $O$.
We consider only density-density terms here, but additional Hund's rule terms would not spoil our reasoning.
We now transform Eq.~(\ref{eqn:Hint}) to the reciprocal space. If one chooses to work in the same basis as in Eq.~(\ref{eqn:1pHam}), one obtains
\begin{align} \notag
 H_\mathrm{int} & = \frac{U_d}{2} \sum_{\sigma,\tau} \int_\mathbb{T} \! d \mathbf{k}_1 \dots d\mathbf{k}_4 \,\, \delta_{\{ \mathbf{k} \}} \,\, d^\dagger_\sigma (\mathbf{k}_1) \, d^\dagger_\tau (\mathbf{k}_2) \, d_\tau (\mathbf{k}_3) \, d_\sigma (\mathbf{k}_4) \\ \notag
 & \quad + \frac{U_p}{2} \sum_{\sigma,\tau,\nu} \int_\mathbb{T} \! d \mathbf{k}_1 \dots d\mathbf{k}_4 \,\, \delta_{\{ \mathbf{k} \}} \,
  (-1)^{(\mathbf{k}_1 + \mathbf{k}_2 - \mathbf{k}_3 - \mathbf{k}_4 ) \cdot \hat{\bm{\nu}} / (2 \pi)} \,\,
 p^\dagger_{\nu,\sigma} (\mathbf{k}_1) \,
p^\dagger_{\nu,\tau} (\mathbf{k}_2)\, p_{\nu,\tau} (\mathbf{k}_3) \, p_{\nu,\sigma} (\mathbf{k}_4) \\ \label{eqn:emery-Hintk}
 & \quad + 2 U_{pd} \sum_{\sigma,\tau,\nu} \int_\mathbb{T} \! d \mathbf{k}_1 \dots d\mathbf{k}_4 \,\, \delta_{\{ \mathbf{k} \}} \, \cos \left[ \frac{(\mathbf{k}_4-\mathbf{k}_1) \cdot
\hat{\bm{\nu}}}{2} \right]
\,\, p^\dagger_{\nu,\sigma} (\mathbf{k}_1) \, d^\dagger_\tau (\mathbf{k}_2) \, d_\tau (\mathbf{k}_3) \, p_{\nu,\sigma} (\mathbf{k}_4) \\
 & = \sum_{\alpha_1,\dots,\alpha_4} \sum_{\sigma_1,\dots,\sigma_4} {\Psi^{\dagger}}_{\sigma_1}^{\alpha_1} (\mathbf{k}_1) \, {\Psi^{\dagger}}_{\sigma_2}^{\alpha_2} (\mathbf{k}_2) \,
 \Psi_{\sigma_3}^{\alpha_3} (\mathbf{k}_3) \, \Psi_{\sigma_4}^{\alpha_4} (\mathbf{k}_4) \,
 \mathcal{V}^{\alpha_1, \dots, \alpha_4}_{\sigma_1, \dots \sigma_4} (\mathbf{k}_1,\mathbf{k}_2,\mathbf{k}_3,\mathbf{k}_4)
\end{align} 
 where
\begin{equation}
 \delta_{\{\mathbf{k} \}} = \left\{ \begin{array}{cl}
 1 & \text{for} \quad (\mathbf{k}_1+\mathbf{k}_2 - \mathbf{k}_3 - \mathbf{k}_4 ) = 2 \pi \mathbf{n} \, , \quad \mathbf{n} \in \mathbb{Z}^2  \\
 0 & \text{otherwise}
\end{array} \right.
\end{equation} 
ensures momentum conservation.
 For an umklapp processes, i.e.\ for $\mathbf{n} \neq 0$, the sign structure of the $U_p$ term is nontrivial due to the improper Bloch basis chosen. The integrals in Eq.~(\ref{eqn:emery-Hintk}) restrict all four momenta to the first BZ. If
 $\mathbf{k}_1 +\mathbf{k}_2 - \mathbf{k}_3$ is related to $\mathbf{k}_4 $ by a non-vanishing reciprocal vector, this term may acquire a minus sign depending
on whether $ n_\nu $ is even or odd.
 Working in an improper Bloch basis requires therefore some care, since ignoring these phases could be pernicious.

 Before we switch to the band language, let us again look at the quadratic part of the Hamiltonian. From Eq.~(\ref{eqn:1pHam}), one can
 observe that $ \mathcal{H}_0( \mathbf{k}) $ does not transform trivially under operations in the point-group of the
underlying square lattice. More precisely, one has $ \mathcal{H}_0( R_{\hat{O}} \mathbf{k}) \neq \mathcal{H}_0( \mathbf{k}) $ for a general point-group operation $ \hat{O} \in C_{4v} $ with a corresponding rotation matrix $ R_{\hat{O}} $ for the  momentum quantum number. Apparently, if the electronic orbitals transform non-trivially under point-group
operations, this gives rise to a tight-binding model with vertex functions that also lack such a trivial behavior. In the present case, the phase of the electronic one-particle wavefunctions transform nontrivially under the point group, as visible in Fig.~\ref{fig:CuO-plane}. This sign structure is inherited by the hopping integrals between these orbitals. For example, hopping from a Copper atom to the oxygen atom right below is inequivalent to hopping to the oxygen atom right above.
 Note that one should avoid speaking of a spontaneously broken symmetry in this case, since
  physical symmetry breaking should rather be associated with certain properties of observables than of auxiliary quantities such as wavefunctions. In our case, however, the occupation of the two $p$-orbitals is equal unless additional terms are included or their degeneracy is broken explicitly, hence the groundstate and responses do not break the point group symmetry.  Therefore,
as will be pointed out in the following, the $C_{4v}$ symmetry is still manifest in the tight-binding Hamiltonian and therefore all observables respect this symmetry, while the phase of the electronic wave function is only an auxiliary quantity. These issues are of course very similar in other fields of physics, but we discuss them here in order to disentangle the various levels. Furthermore, the discussion beyond the one-particle level is usually not undertaken.

In this place, the following questions seem appropriate:
\begin{enumerate}[i)]
 \item In what way is the point-group symmetry of the lattice manifest in the Emery model? \label{enu:remainder}
 \item Is there an alternative, explicitly $C_{4v}$-symmetric formulation of the Emery model with vertex functions that behave trivially under the point-group operations?
\end{enumerate}
Let us now look for how the point group symmetry manifests itself formally.
\subsection{$C_{4v} $ symmetry} \label{sec:hidden-c4v}

\begin{figure}
 \centering
 \includegraphics[width=0.26\linewidth]{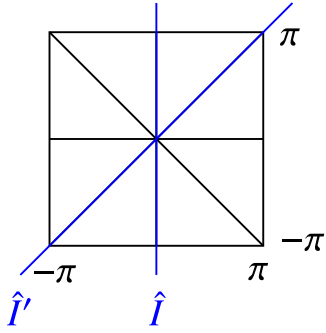}
\caption{Symmetry elements of $ C_{4v} $ in the first BZ $ \mathbb{T} $ of the Emery model. All $C_{4v} $ operations can be interpreted as products of mirror operations
$ \hat{I} $ and $ \hat{I}' $ with respect to the blue axes.} \label{fig:c4v-symmel}
\end{figure}

In the (improper) Bloch basis of Eq.~(\ref{eqn:1pHam}), in addition to a real-valued coupling function  $\mathcal{H}_0 (\mathbf{k}) $, the Hamiltonian shows a nice behavior under point-group operations $ \hat{O} \in C_{4v} $. These operations can always be written as a product of the mirror operations $ \hat{I} $ and $ \hat{I}' $ with respect to the $ y $ axis and a BZ diagonal, respectively. More precisely, we define $ \hat{I'} $ as the permutation operation on the coordinates $ R_{\hat{I}'} \mathbf{k} = (k_y,k_x)^\mathrm{T} $.
In Fig.~\ref{fig:c4v-symmel}, one can easily see that the other symmetry elements of $ C_{4v} $ can be generated by successive application of $ \hat{I} $ and $\hat{I}'$.
Under the reflection  $ \hat{I} $ of the $x$ coordinate, the one-particle coupling function transforms as
\begin{equation} \label{eqn:Emery-I}
 \mathcal{H}_0 (R_{\hat{I}} \mathbf{k}) =  M_{\hat{I}} \, \mathcal{H}_0 (\mathbf{k}) \, M^\dagger_{\hat{I}} \, , \quad
 M_{\hat{I}} =  \left( \begin{array}{rrr} 1 & 0 &0\\ 0 & -1 & 0 \\ 0 & 0 & 1 \end{array} \right) \, ,
\end{equation} 
i.e.\ the sign of the hybridization matrix element between the $ d$- and $p_x$-orbitals gets flipped. This is due to the non-trivial $C_{4v} $-behavior of specific orbitals.

Under coordinate exchange $ \hat{I}' $, $ \mathcal{H}_0 (\mathbf{k}) $ also shows a simple behavior. The $p_x $- and $ p_y$-orbitals then change their roles and we have
\begin{equation} \label{eqn:Emery-I'}
 \mathcal{H}_0 (R_{\hat{I}'} \mathbf{k}) =  M_{\hat{I}'} \, \mathcal{H}_0 (\mathbf{k}) \, M^\dagger_{\hat{I}'} \, , \quad
 M_{\hat{I}'} =  \left( \begin{array}{rrr} 1 & 0 &0\\ 0 & 0 & 1 \\ 0 & 1 & 0 \end{array} \right) \, .
\end{equation} 
Note that this property does not arise from a non-trivial point-group behavior of the electronic structure, but from the presence of
 two atoms of the same kind in different locations in the unit cell.
 Since, in the present example, $ \hat{I} '$ maps these two atoms onto one another, one should not expect $ \mathcal{H}_0 (\mathbf{k}) $ to transform trivially under $ \hat{I}'$.
So altogether, we have found that the point-group symmetry of the lattice is hidden in Eqs.~(\ref{eqn:Emery-I}) and (\ref{eqn:Emery-I'}). In other words, the one-particle Hamiltonian
$ H_0 $ is invariant under the transformations
\begin{equation} \label{eqn:transf-emeryI}
 \Psi_\sigma (\mathbf{k} ) \to \Psi'_\sigma (R_{\hat{I}} \mathbf{k} ) = M_{\hat{I}} \Psi_\sigma (\mathbf{k}) \, , \quad
 \mathcal{H}_0 (\mathbf{k}) \to \mathcal{H}_0 (R_{\hat{I}} \mathbf{k})
\end{equation} 
 and
\begin{equation} \label{eqn:transf-emeryIp}
 \Psi_\sigma (\mathbf{k} ) \to \Psi'_\sigma (R_{\hat{I}'} \mathbf{k} ) = M_{\hat{I}'} \Psi_\sigma (\mathbf{k}) \, , \quad
 \mathcal{H}_0 (\mathbf{k}) \to \mathcal{H}_0 (R_{\hat{I}'} \mathbf{k})
\end{equation} 
in the improper Bloch basis of Eq.~(\ref{eqn:1pHam}).

 As will become clear in Sec.~\ref{sec:trans-rule}, the matrices $ M_{\hat{I}} $ and $ M_{\hat{I}'} $ directly stem from the Bloch orbitals in the corresponding basis according to
\begin{equation}
 \left( M_{\hat{O}} \right)_{\alpha,\beta} = \int \! d \mathbf{r} \, \left\langle R_{\hat{O}} \mathbf{r} \left| \phi_\sigma^\alpha (R_{\hat{O}} \mathbf{k}) \right. \right\rangle
\left\langle \left. \phi_\sigma^\beta (\mathbf{k}) \right| \mathbf{r} \right\rangle
\end{equation} 
 for arbitrary spin orientation $\sigma $ and $ \hat{O} \in C_{4v} $. The rotated position $ R_{\hat{O}} \mathbf{r} $ in the first scalar product probes the symmetry of the
electronic wavefunctions, giving rise to the precise form of $M_{\hat{I}} $.
 In addition, it accounts for the action of $\hat{O} $ on the nuclear positions, since some of the corresponding Wannier states $ \left| \psi_\sigma^\alpha (\mathbf{R}) \right\rangle $
 may belong to atoms away form the center $ \mathbf{R} $ of the respective unit cell according to the conventions introduced in Sec.~\ref{sec:Wannier}.
 In the present example, this leads to the representation matrix $ M_{\hat{I}'} $.
 Moreover, the wavevector $ \mathbf{k} $ is rotated to $ R_{\hat{O}} \mathbf{k} $ in the first scalar product, which corresponds to a rotation of the direct unit cells,
 $\mathbf{k} $ being the reciprocal space variable corresponding to their centers $\mathbf{R}$.

 These representation matrices are hence not fully determined by the point-group behavior of the electronic orbitals, but also the point-group behavior of the nuclear
 positions matters.
 In this context, we would like to recall that the electronic orbitals correspond to basis functions of the irreducible representations of $ D_{4h} $, since they are truly three-dimensional.
 As a lattice model of a CuO plane, the Emery model is however only two-dimensional and therefore the point group is reduced to $ C_{4v} $.
 For the Emery model, the representation matrices $ M_{\hat{I}} $ and $ M_{\hat{I}'} $ (and consequently of all other operations $ \hat{O} \in C_{4v} $) decay into irreducible blocks
 -- for the Cu orbitals transforming with the irreducible
 representation $A_1$ and for the O $p$-orbitals transforming with $E$. The reader should be aware that this is due to the lattice structure. For the example of graphene
 in the following section, reducible representation matrices will emerge from irreducible electronic orbitals centered around inequivalent lattice positions.

 Let us now return to the transformation behavior of the Emery Hamiltonian.
 The one-particle Hamiltonian density
\begin{equation}
 \Psi^\dagger_\sigma (\mathbf{k}) \, \mathcal{H}_0 (\mathbf{k}) \,\Psi_\sigma (\mathbf{k})
\end{equation} 
transforms to
\begin{equation}
 {\Psi'}^\dagger_\sigma (R_{\hat{O}} \mathbf{k}) \, \mathcal{H}_0 (R_{\hat{O}} \mathbf{k}) \,{\Psi'}_\sigma (R_{\hat{O}} \mathbf{k})  =
 \Psi^\dagger_\sigma (\mathbf{k}) \, M^\dagger_{\hat{O}} \, \mathcal{H}_0 (R_{\hat{O}} \mathbf{k}) \,M_{\hat{O}} \, \Psi_\sigma (\mathbf{k}) \, , \quad \hat{O} = \hat{I},\hat{I}' \, .
\end{equation} 
 The invariance of the one-particle Hamiltonian density then follows from
\begin{equation}
 \mathcal{H}_0 (R_{\hat{O}} \mathbf{k}) = M_{\hat{O}} \, \mathcal{H}_0 (\mathbf{k}) \, M^\dagger_{\hat{O}}
\, , \quad \hat{O} = \hat{I},\hat{I}' \, ,
\end{equation} 
i.e., it looks the same in the original and in the transformed frame.

Let us check this invariance claim also for the interacting part $ H_\mathrm{int} $ in Eq.~(\ref{eqn:emery-Hintk}) of the Hamiltonian.
 As one may easily verify, it is invariant under the transformations in Eqs.~(\ref{eqn:transf-emeryI}) and (\ref{eqn:transf-emeryIp}) and therefore the
full, interacting Emery Hamiltonian has a manifest $ C_{4v} $ symmetry.
 More formally, we have
\begin{align} \notag
  & \mathcal{V}_{a_1, \dots, a_{2n}}^{\alpha_1,\dots,\alpha_{2n}}  ( \mathbf{k}_1, \dots, \mathbf{k}_n; \mathbf{k}_{n+1}, \dots, \mathbf{k}_{2n}) \,\,
 {\Psi_{a_1}^{\alpha_1}}^\dagger (\mathbf{k}_1) \dots {\Psi_{a_n}^{\alpha_n}}^\dagger (\mathbf{k}_n) \,\,
 {\Psi_{a_{n+1}}^{\alpha_{n+1}}}(\mathbf{k}_{n+1}) \dots {\Psi_{a_{2n}}^{\alpha_{2n}}} (\mathbf{k}_{2n}) \\ \notag
 \to \quad & \mathcal{V}_{a_1, \dots, a_{2n}}^{\alpha_1,\dots,\alpha_{2n}}  (R_{\hat{O}} \mathbf{k}_1, \dots, R_{\hat{O}} \mathbf{k}_{2n}) \,\,
 {{\Psi'}_{a_1}^{\alpha_1}}^\dagger (R_{\hat{O}} \mathbf{k}_1) \dots {{\Psi'}_{a_n}^{\alpha_n}}^\dagger (R_{\hat{O}} \mathbf{k}_n) \\
 & \quad \times {{\Psi'}_{a_{n+1}}^{\alpha_{n+1}}}(R_{\hat{O}} \mathbf{k}_{n+1}) \dots {{\Psi'}_{a_{2n}}^{\alpha_{2n}}} (R_{\hat{O}} \mathbf{k}_{2n})
 \, ,
\end{align} 
 where
\begin{equation}
 \mathcal{V}_{\bm{\sigma}}^{\alpha_1 \dots \alpha_4} (R_{\hat{O}} \mathbf{k}_1, \dots, R_{\hat{O}}\mathbf{k}_4)
 = {M_{\hat{O}}}_{\alpha_1,\beta_1} {M_{\hat{O}}}_{\alpha_2,\beta_2} \mathcal{V}_{\bm{\sigma}}^{\beta_1 \dots \beta_4} (\mathbf{k}_1, \dots, \mathbf{k}_4) \, {M^\dagger_{\hat{O}}}_{\alpha_3,\beta_3} {M^\dagger_{\hat{O}}}_{\alpha_4,\beta_4}
\end{equation} 
 for $ \hat{O} \in C_{4v} $, i.e. the interaction function has the same components of the coupling function in the transformed frame as in the original one. However, this equality requires to transform the electron operators according to Eqs.~(\ref{eqn:transf-emeryI}) and (\ref{eqn:transf-emeryIp}). In the following subsection, we will show that this somewhat hidden symmetry translates to a more explicit one in the band language for a suitably chosen band gauge.
\subsection{Band language and natural Bloch basis} \label{sec:Emery-natural}

 As in Eq.~(\ref{eqn:intro-u}), let us now switch to new fields $ \bm{\chi}_\sigma (\mathbf{k}) = u (\mathbf{k}) \, \Psi_\sigma (\mathbf{k} ) $ with wavevector-dependent, unitary $ u (\mathbf{k}) $ in which
the one-particle Hamiltonian
\begin{equation}
  H_0  = \sum_{\sigma} \int_\mathbb{T} \! d\mathbf{k} \,  \Psi_\sigma^\dagger (\mathbf{k}) \, \mathcal{H}_0 (\mathbf{k}) \, \Psi_\sigma (\mathbf{k})
  = \sum_{\sigma} \int_\mathbb{T} \! d\mathbf{k} \,  \bm{\chi}_\sigma^\dagger (\mathbf{k}) \, \mathcal{B}_0 (\mathbf{k}) \, \bm{\chi}_\sigma (\mathbf{k})
\end{equation} 
 is diagonal. The diagonalized coupling function
\begin{equation}
\mathcal{B} (\mathbf{k}) = u (\mathbf{k}) \, \mathcal{H}_0 (\mathbf{k}) \, u^\dagger (\mathbf{k})
 = \left( \begin{array}{ccc}
 \frac{\epsilon}{2} [1 - r (\mathbf{k})] & 0 &0\\
 0 & \epsilon &0 \\
 0 & 0& \frac{\epsilon}{2} [1 + r (\mathbf{k}) ]
\end{array} \right)
\end{equation} 
then contains the band dispersion with the short-hand notation
\begin{equation}
 r (\mathbf{k}) = \sqrt{ 1 + 16 \left(\frac{t}{\epsilon} \right)^2 \left[ \sin^2 (k_x/2) + \sin^2 (k_y/2) \right]} \, .
\end{equation} 
 We have labeled the bands according to the energies in ascending order.
 In this order, they correspond to the antibonding, nonbonding and bonding solutions of the quadratic part of the Hamiltonian.
 One possible choice for the transformation matrices $ u (\mathbf{k} )$ then reads as
\begin{equation} \label{eqn:u-emery}
 u (\mathbf{k}) = \left( \begin{array}{ccc}
 \frac{\epsilon}{4t} N_1 (\mathbf{k}) [ 1 + r (\mathbf{k})] & - \sin (k_x/2)\, N_1 (\mathbf{k})  &- \sin (k_y/2)\, N_1 (\mathbf{k}) \\
 0 & - \sin (k_y/2)\, N_2 (\mathbf{k}) & \sin (k_x/2) \, N_2 (\mathbf{k}) \\
 \frac{\epsilon}{4t} N_3 (\mathbf{k}) [ 1 - r (\mathbf{k})] & - \sin (k_x/2)\, N_3 (\mathbf{k})  & -\sin (k_y/2)\, N_3 (\mathbf{k})
 \end{array} \right)
\end{equation} 
with normalization factors
\begin{align}
 N_1 (\mathbf{k}) &= \left\{ \frac{\epsilon}{4t} \left[1+ r(\mathbf{k}) \right]^2 +
\sin^2 (k_x/2) + \sin^2 (k_y/2) \right\}^{-1/2} \\
 N_2 (\mathbf{k}) &= \left[ \sin^2 (k_x/2) + \sin^2 (k_y/2) \right]^{-1/2} \\
 N_3 (\mathbf{k}) &= \left\{ \frac{\epsilon}{4t} \left[1- r(\mathbf{k}) \right]^2 +
\sin^2 (k_x/2) + \sin^2 (k_y/2) \right\}^{-1/2}  \, .
\end{align} 
Note that this transformation matrix inherits discontinuities at the boundary of the BZ, since we have started from an improper basis of the non-hybridizing Bloch states.
Moreover, the bands with labels $2$ and $3$ are degenerate at $ \mathbf{k} = 0 $ and so there is some freedom in choosing $ u (0) $. One possibility results from the limit
\begin{equation}
 \lim_{k_y \to 0} u ( 0, k_y) = \left( \begin{array}{rrr}
 1 & 0 & 0 \\
 0 & -1 & 0 \\
 0 & 0 & -1
 \end{array} \right)
\end{equation} 
while another one is obtained by approaching the origin on the $y$ axis, i.e.
\begin{equation}
 \lim_{k_x \to 0} u ( k_x, 0) = \left( \begin{array}{rrr}
 1 & 0 & 0 \\
 0 & 0 & 1 \\
 0 & 1 & 0
 \end{array} \right) \, .
\end{equation} 
Therefore, there must a discontinuity at the origin. In contrast to the discontinuities at the BZ boundary, it is not lifted in
the proper Bloch basis. Moreover, one can observe that the lowest band only has $d$-wave character at the origin, while the other two ones are degenerate and consist purely of the $p$-orbitals.
 We will further elaborate on
 such discontinuities occurring on symmetry elements (mirror axes or planes or inversion centers, for example) of the point group when we discuss the general case.

 In a next step, we look at the interaction in the band language. It reads as
\begin{equation}
 H_\mathrm{int} = \sum_{\sigma_1 \dots \sigma_4} \sum_{\alpha_1 \dots \alpha_4} \int_\mathbb{T} \! d \mathbf{k}_1 \dots d \mathbf{k}_4 \,\,\delta_{\{ \mathbf{k} \}} \, \delta_{\sigma_1,\sigma_2} \,\delta_{\sigma_3,\sigma_4} \, f^{\bm{\alpha}} (\mathbf{k}_1,\mathbf{k}_2,\mathbf{k}_3,\mathbf{k}_4) \,
 \chi^\dagger (\xi_1)\, \chi^\dagger (\xi_2) \, \chi (\xi_3) \, \chi (\xi_4) \, ,
\end{equation} 
 with the coupling function
 \begin{align} \notag
 f^{\bm{\alpha}} (\mathbf{k}_1,\mathbf{k}_2,\mathbf{k}_3,\mathbf{k}_4) & = \frac{U_d}{2} \, u_{\alpha_1,1} (\mathbf{k}_1) \, u_{\alpha_2,1} (\mathbf{k}_2) \, u_{\alpha_3,1}^\ast (\mathbf{k}_3) \, u_{\alpha_4,1}^\ast (\mathbf{k}_4) \\ \notag
 & \quad + \sum_\nu \frac{U_p}{2} \,
  (-1)^{(\mathbf{k}_1 + \mathbf{k}_2 - \mathbf{k}_3 - \mathbf{k}_4 ) \cdot \hat{\bm{\nu}} / (2 \pi)} \,\,
 u_{\alpha_1,\nu} (\mathbf{k}_1) \,  u_{\alpha_2,\nu} (\mathbf{k}_2) \,
u_{\alpha_3,\nu}^\ast (\mathbf{k}_3)\, u_{\alpha_4,\nu}^\ast (\mathbf{k}_4) \\ \label{eqn:int-band}
 & \quad + \sum_\nu 2 U_{pd} \, \cos \left[ \frac{(\mathbf{k}_4-\mathbf{k}_1) \cdot
\hat{\bm{\nu}}}{2} \right]
\,\, u_{\alpha_1,\nu} (\mathbf{k}_1) \, u_{\alpha_2,1} (\mathbf{k}_2) \, u_{\alpha_3,1}^\ast (\mathbf{k}_3) \, u_{\alpha_4,\nu}^\ast (\mathbf{k}_4) \, ,
 \end{align} 
 where $ \xi_i = (\alpha_i,\sigma_i,\mathbf{k}_i) $ includes the band index $\alpha_i$ as well as spin and momentum quantum numbers.
 Even the $U_d$ term, which has a trivial momentum dependence in the orbital language, now acquires some orbital makeup through the transformation $u(\mathbf{k}) $.

Let us now discuss the point-group behavior in the band language. One can observe that the band dispersion behaves trivially under point-group operations, i.e., that $ \mathcal{B} (R_{\hat{O}} \mathbf{k} ) = \mathcal{B} (\mathbf{k}) $. The question now is whether this also holds for the vertex function of the two-particle interaction.
The transformation matrix elements $u (\mathbf{k}) $ in Eq.~(\ref{eqn:int-band}) transform trivially for the $d$-orbital irrespective of the band index $\alpha$, i.e.
\begin{equation}
 u_{\alpha,1} (R_{\hat{O}} \mathbf{k}) = u_{\alpha,1} (\mathbf{k}) \quad \forall \hat{O} \in C_{4v} \, ,
\end{equation} 
while the matrix elements for the $p$-orbitals show a non-trivial behavior. If these other elements were to transform according to
\begin{equation} \label{eqn:Icons-em}
 u_{\alpha,2} (R_{\hat{I}} \mathbf{k}) = - u_{\alpha,2} (\mathbf{k})
\end{equation} 
and
\begin{equation} \label{eqn:I'cons-em}
 u_{\alpha,2} (R_{\hat{I}'} \mathbf{k}) =  u_{\alpha,3} (\mathbf{k}) \, , \quad
 u_{\alpha,3} (R_{\hat{I}'} \mathbf{k}) =  u_{\alpha,2} (\mathbf{k}) \, ,
\end{equation} 
 the coupling function $f$ would be invariant under $ \mathbf{k}_i \to R_{\hat{O}} \mathbf{k}_i $ $ \forall \hat{O} \in C_{4v} $.
 These conditions follow from the form of $f$ in Eq.~(\ref{eqn:int-band}) up to the signs, which are fixed by the quadratic part given in Eq.~(\ref{eqn:1pHam}).
 A fourth (redundant) condition
\begin{equation}
 u_{\alpha,3} (R_{\hat{J}} \mathbf{k}) = - u_{\alpha,3} (\mathbf{k}) \, , \quad \hat{J} = \hat{I}'\hat{I}\hat{I}'
\end{equation} 
 follows from Eqs.~(\ref{eqn:Icons-em}) and (\ref{eqn:I'cons-em}).

 For $\alpha =2$, however, the transformation matrix given in Eq.~(\ref{eqn:u-emery}) violates Eqs.~(\ref{eqn:Icons-em}) and (\ref{eqn:I'cons-em}). But there is some freedom in the choice of $ u (\mathbf{k}) $, as eigenvectors of complex matrices are only determined up
 to a phase factor. In the present case, changing these phase factors corresponds to a basis transformation between Bloch states. Moreover, this phase can be fixed locally in momentum space
 or, in other words, individually for each $\mathbf{k}$.  This way, one may introduce additional discontinuities in the transformation matrix. This is also the case for non-hybridizing Bloch states
 with $ u (\mathbf{k}) $ according to Eq.~(\ref{eqn:u-emery}) for $ k_y \geq k_x \geq 0 $ or $ k_y < k_x < 0$ and
\begin{equation}
 u (\mathbf{k}) = \left( \begin{array}{ccc}
 \frac{\epsilon}{4t} N_1 (\mathbf{k}) [ 1 + r (\mathbf{k})] & - \sin (k_x/2)\, N_1 (\mathbf{k})  &- \sin (k_y/2)\, N_1 (\mathbf{k}) \\
 0 &  \sin (k_y/2)\, N_2 (\mathbf{k}) & - \sin (k_x/2) \, N_2 (\mathbf{k}) \\
 \frac{\epsilon}{4t} N_3 (\mathbf{k}) [ 1 - r (\mathbf{k})] & - \sin (k_x/2)\, N_3 (\mathbf{k})  & -\sin (k_y/2)\, N_3 (\mathbf{k})
 \end{array} \right)
\end{equation} 
 otherwise. In other words, the flat band with index $ \alpha=2$ is multiplied by a factor $ -1$  for $ k_x \to - k_x $ and for $ k_x  \leftrightarrow k_y $. We observe, that Eqs.~(\ref{eqn:Icons-em}) and (\ref{eqn:I'cons-em}) are fulfilled
 in this new basis and that $f$ hence behaves trivially under all $C_{4v}$ operations, i.e.
\begin{equation}
 f^{\bm{\alpha}} (R_{\hat{O}} \mathbf{k}_1,R_{\hat{O}} \mathbf{k}_2,R_{\hat{O}} \mathbf{k}_3,R_{\hat{O}} \mathbf{k}_4)
 = f^{\bm{\alpha}} ( \mathbf{k}_1, \mathbf{k}_2, \mathbf{k}_3, \mathbf{k}_4) \quad \forall\, \hat{O} \in C_{4v} \, .
\end{equation} 
This implies, that the full Hamiltonian is symmetric under
\begin{equation} \label{eqn:expl-em}
 \bm{\chi}_\sigma (\mathbf{k}) \to \bm{\chi'}_\sigma (R_{\hat{O}} \mathbf{k}) = \bm{\chi}_\sigma ( \mathbf{k}) \quad \forall \hat{O} \in C_{4v} \, ,
\end{equation} 
i.e.\ the electron band operators do not need to be transformed or reordered.
Due to this explicit symmetry, we call the new Bloch basis a \emph{natural} one. In a functional integral language, the invariance of the Hamiltonian under the transformation in Eq.~(\ref{eqn:expl-em}) translates to a trivial point-group behavior of all coupling functions of the action and the generating functionals of amputated Green's function and one-particle irreducible vertices. This way, the point-group symmetry can be exploited straightforwardly in a fRG approach.
\begin{figure}
 \centering
 \includegraphics[width=0.5\linewidth]{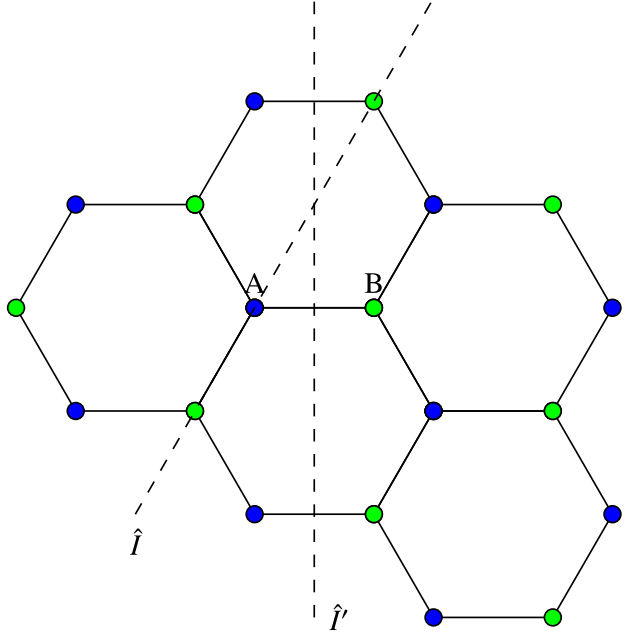}
\caption{Honeycomb lattice with interpenetrating sublattices $A$ and $B$. Also the mirror axes of the point-group operations $ \hat{I} $ and $ \hat{I}'$ are depicted here.} \label{fig:honeyc-lattice}
\end{figure}
\section{Sixfold symmetry: Graphene} \label{sec:graphene}

 \subsection{Model Hamiltonian} \label{sec:hidden-c6v}

  Before we discuss the general case, also an example of a six-fold symmetry shall be given. We consider a model for spinful fermions on the honeycomb lattice describing the $p_z$-orbitals in a
graphene monolayer (for a review see Ref.~\onlinecite{graphene-rmp}). Such a tight-binding description of graphene has a long history.  \cite{graphene-wallace} Also other materials with a honeycomb lattice, such as $ \mathrm{In}_3 \mathrm{Cu}_2 \mathrm{VO}_9 $ have been studied recently.  \cite{kataev-incu,moeller-incu,yan-incu} As can be seen from Fig.~\ref{fig:honeyc-lattice}, we are dealing with two interpenetrating sublattices with creation operators $ a^\dagger_{\sigma} (\mathbf{R}) $ and $ b^\dagger_{\sigma} (\mathbf{R}) $, where  $\mathbf{R}$ denotes the position of the unit cell. In the following, we assign the position quantum number $\mathbf{R}$ to an $A$ site and the $B$ site which is the nearest neighbor to its right.
The two other nearest neighbors of the $A$ site then are attributed to the unit cells with $\mathbf{R} + \delta_2 $ and $ \mathbf{R}+\delta_3$. If the distance between nearest neighbors is again set to unity, one has
 $ \delta_1 =0 $, $ \delta_2 = ( -3/2, \sqrt{3} /2 ) $ and $ \delta_3 =  ( -3/2, -\sqrt{3} /2 ) $ for the primitive translations.
 In the following, we consider a Hamiltonian
\begin{align} \notag
 H & = -t \sum_{\sigma,\mathbf{R},\delta} \left[ a^\dagger_\sigma (\mathbf{R}) \, b_\sigma (\mathbf{R}+\delta) + b^\dagger_\sigma (\mathbf{R}+\delta) \, a_\sigma (\mathbf{R}) \right] \\ \notag
 & \quad + U \sum_\mathbf{R} : n^{(a)}_\uparrow (\mathbf{R}) \, n^{(a)}_\downarrow  (\mathbf{R}) +  n^{(b)}_\uparrow (\mathbf{R}) \, n^{(b)}_\downarrow (\mathbf{R}) : \\ \label{eqn:graphene}
 & \quad + V \sum_{\mathbf{R},\delta,\sigma,\tau}  :n^{(a)}_\sigma (\mathbf{R}) \, n^{(b)}_\tau  (\mathbf{R}+\delta) + n^{(b)}_\sigma (\mathbf{R}+\delta) \, n^{(a)}_\tau (\mathbf{R}) :
\end{align} 
with nearest neighbor hopping $t$, on-site interaction $U$ and nearest-neighbor interaction $V$.
 \begin{figure}
 \centering
 \includegraphics[width=0.26\linewidth]{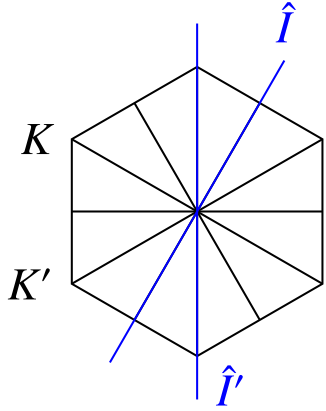}
\caption{First BZ $\mathbb{H} $ for the honeycomb lattice  with lines corresponding to mirror axes. All operations in $C_{6v} $ can be written as products of the mirror operations with
axes $ \hat{I} $ and $ \hat{I}'$.}\label{fig:honeyc-BZ}
\end{figure}
As before, we first look at the quadratic part $H_0$ and switch to reciprocal space according to
\begin{align}
 a_\sigma(\mathbf{R}) & = \int_\mathbb{H} \! d\mathbf{k} \, e^{i \mathbf{k} \cdot \mathbf{R}} \, a_\sigma (\mathbf{k}) \\
 b_\sigma (\mathbf{R}) & = \int_\mathbb{H} \! d\mathbf{k} \, e^{i \mathbf{k} \cdot \mathbf{R}} \, b_\sigma (\mathbf{k})
\end{align} 
where the momentum integrals run over the first BZ $\mathbb{H} $ depicted in Fig.~\ref{fig:honeyc-BZ}. In terms of the orbitors
\begin{equation} \label{eqn:graphene-proper}
 \Psi_\sigma (\mathbf{k}) = \left( \begin{array}{c} a_\sigma (\mathbf{k}) \\ b_\sigma (\mathbf{k}) \end{array} \right) \, ,
\end{equation} 
the one-particle Hamiltonian reads
\begin{equation}
 H_0 = -t \sum_\sigma \int_\mathbb{H} \! d\mathbf{k} \, \Psi^\dagger_\sigma (\mathbf{k}) \, \mathcal{H}_0 (\mathbf{k}) \, \Psi_\sigma (\mathbf{k}) \, ,
\quad
 \mathcal{H}_0 (\mathbf{k}) = \left( \begin{array}{cc} 0 & h (\mathbf{k}) \\ h (\mathbf{k})^\ast & 0 \end{array} \right)
\end{equation} 
with $ h (\mathbf{k}) = \sum_\delta e^{i \delta \cdot \mathbf{k}} $. The one-particle coupling function $ \mathcal{H}_0 (\mathbf{k})  $
 vanishes at $ K = 2 \pi (-1/\sqrt{3},1/3)$ and $K' =  2 \pi (-1/\sqrt{3},-1/3)$. If $ \mathcal{H}_0 (\mathbf{k})  $ is expanded around these points, one obtains
 a Dirac Hamiltonian.  \cite{graphite1958}

 In reciprocal space, the interacting part $ H_\mathrm{int} $ of the Hamiltonian reads as
\begin{align} \notag
 H_\mathrm{int} & = U \int_\mathbb{H} \! d\mathbf{k}_1 \dots d\mathbf{k}_4 \, \delta_{\{ \mathbf{k} \}}
\left[ a^\dagger_\uparrow (\mathbf{k}_1) \,a^\dagger_\downarrow (\mathbf{k}_2) \, a_\downarrow (\mathbf{k}_3) \,a_\uparrow (\mathbf{k}_4) \,
 + b^\dagger_\uparrow (\mathbf{k}_1) \,b^\dagger_\downarrow (\mathbf{k}_2) \, b_\downarrow (\mathbf{k}_3) \,b_\uparrow (\mathbf{k}_4) \right] \\ \notag
 & \quad + V \sum_{\sigma,\tau} \int_\mathbb{H} \! d\mathbf{k}_1 \dots d\mathbf{k}_4 \, \delta_{\{ \mathbf{k} \}} \left[ h (\mathbf{k}_3 - \mathbf{k}_2 ) \,\, a^\dagger_\sigma (\mathbf{k}_1) \,b^\dagger_\tau (\mathbf{k}_2) \,b_\tau (\mathbf{k}_3) \, a_\sigma (\mathbf{k}_4)  \right. \\
 & \qquad + \left. h (\mathbf{k}_4 - \mathbf{k}_1 ) \,\, b^\dagger_\sigma (\mathbf{k}_1) \, a^\dagger_\tau (\mathbf{k}_2) \,a_\tau (\mathbf{k}_3) \,b_\sigma (\mathbf{k}_4)  \right] \, ,
\end{align} 
where $ \delta_{\{ \mathbf{k}\}} $ again ensures momentum conservation up to reciprocal lattice vectors.

 As the spacings $ \delta $ are primitive vectors of the direct lattice, $ h (\mathbf{k}) $ is periodic in reciprocal space or, if all momenta are folded back to the
first BZ, continuous at the zone boundary. We are hence already working in a proper basis. For the present example, the behavior under point-group operations is already fairly simple in this basis:
 In the case of the honeycomb lattice, the point group is $ C_{6v} $. All operations of this group can be written as products of  two mirror operations
$ \hat{I} $ and $ \hat{I}'$
 with respect to axes going through the lattice sites and the middle of the bonds between neighboring sites, respectively. In Figs.~\ref{fig:honeyc-lattice} and \ref{fig:honeyc-BZ},
 we have chosen the $y$ axis to coincide with the mirror axis of $ \hat{I}' $, while the mirror axis of $ \hat{I}$ is rotated by $ \pi /6 $ with respect to $ y $ axis.
 Clearly, $\hat{I}$ maps the two sublattices onto themselves and $\hat{I}'$ maps them onto on another.
 We observe that $ h (R_{\hat{I}} \mathbf{k} ) = h (\mathbf{k}) $ and $ h (R_{\hat{I}'} \mathbf{k} ) = h (\mathbf{k})^\ast $. This leads to a simple behavior of the one-particle
coupling function
\begin{equation}
 \mathcal{H}_0 (R_{\hat{I}} \mathbf{k}) = \mathcal{H}_0 ( \mathbf{k}) \, , \quad
 \mathcal{H}_0 (R_{\hat{I}'} \mathbf{k}) = M_{\hat{I}'} \, \mathcal{H}_0 ( \mathbf{k}) \, M^\dagger_{\hat{I}'} \, , \quad
 M_{\hat{I}'} = \left( \begin{array}{cc} 0 & 1\\ 1 & 0 \end{array} \right) \, .
\end{equation} 
As for
the interacting part, one may substitute $ \delta_{\{ \mathbf{k} \}} h (\mathbf{k}_4 - \mathbf{k}_1) = \delta_{\{ \mathbf{k} \}} h (\mathbf{k}_3 - \mathbf{k}_2)^\ast $.
Therefore,  $ H_\mathrm{int} $ is left unchanged under $ \mathbf{k}_i \to R_{\hat{I}} \mathbf{k}_i $ and the $ a$ and $ b$ operators are interchanged under $ \mathbf{k}_i \to R_{\hat{I}'} \mathbf{k}_i $ .
So altogether, the full Hamiltonian is invariant under the two operations
\begin{equation}
 \Psi_\sigma (\mathbf{k}) \to {\Psi'}_\sigma (R_{\hat{I}} \mathbf{k}) = \Psi_\sigma (\mathbf{k}) \, , \quad
 \Psi_\sigma (\mathbf{k}) \to {\Psi'}_\sigma (R_{\hat{I}'} \mathbf{k}) = M_{\hat{I}'} \Psi_\sigma (\mathbf{k}) \, .
\end{equation} 
This reflects the $ C_{6v} $ symmetry of the system, since all operations in this group can be written as products of the identity, $\hat{I} $ and $\hat{I}'$.

 Note that the resulting representation matrices $M_{\hat{O}} $ are reducible. This may seem counterintuitive in first place, since the underlying (hybridizing) Bloch states
 have $p_z$
 character, which corresponds to the $ A_{2u} $ irreducible representation of $ D_{6h} $. (In a way similar as in the case of the Emery model, the point group $ D_{6h} $ of the
 three-dimensional electronic structure gets reduced to $ C_{6v} $ in the two-dimensional lattice model.) Since one has two inequivalent sites per unit-cell, however, the resulting
 nontrivial point-group behavior of the nuclear positions gives rise to reducible representation matrices.
 In the Bloch basis with states
\begin{equation}
 \left| c^\pm_\sigma  \right\rangle = \frac{1}{\sqrt{2}} \left[ a^\dagger_\sigma (\mathbf{k}) \pm b^\dagger_\sigma (\mathbf{k}) \right]  \left| 0 \right\rangle  \, ,
\end{equation} 
 one would obtain representation matrices with blocks corresponding to the $A_1$ and $B_2$ irreducible representations of $ C_{6v} $, i.e.\
\begin{equation}
 \tilde{M}_{\hat{I}} = \left( \begin{array}{rr} 1 & 0\\ 0 & 1 \end{array} \right) \, , \qquad
 \tilde{M}_{\hat{I}'} = \left( \begin{array}{rr} 1 & 0\\ 0 & -1 \end{array} \right) \, .
\end{equation} 
 Note this is not yet the band basis and therefore the states $  \left| c^\pm_\sigma  \right\rangle $ hybridize almost everywhere on the BZ.
 For the considerations in this work, the reducibility or irreducibility of the representation matrices does not play a role and switching from reducible to irreducible orbitals
 may be of little practical use.
\subsection{Band language and natural basis} \label{sec:graphene-natural}

 Now we again switch to the band language, where one has
\begin{equation}
 H_0 =  \sum_\sigma \int_\mathbb{H} \! d\mathbf{k} \, \bm{\chi}^\dagger_\sigma (\mathbf{k}) \, \mathcal{B} (\mathbf{k}) \, \bm{\chi}_\sigma (\mathbf{k}) \, ,
\quad
 \mathcal{B}(\mathbf{k}) = t \left( \begin{array}{cc}  + \left| h (\mathbf{k}) \right| & 0 \\ 0 &- \left| h (\mathbf{k}) \right|  \end{array} \right)
\end{equation} 
 for the one-particle Hamiltonian. Again the dispersion transforms trivially, i.e.\ $ \mathcal{B} (R_{\hat{O}} \mathbf{k} ) =\mathcal{B} (\mathbf{k} ) $ $ \forall \hat{O} \in C_{6v} $.
The band operators $ \bm{\chi}_\sigma (\mathbf{k}) = u (\mathbf{k}) \, \Psi_\sigma (\mathbf{k})  $ are obtained from the orbitors $ \Psi_\sigma (\mathbf{k}) $ by transformation matrices
\begin{equation} \label{eqn:graphene-trans}
  u (\mathbf{k}) = \frac{1}{\sqrt{2}} \left( \begin{array}{ll} e^{i \phi (\mathbf{k})} & -1 \\ 1 & e^{-i \phi (\mathbf{k})} \end{array} \right) \, .
\end{equation} 
The phase $ \phi (\mathbf{k}) = -i \ln \left[ h(\mathbf{k}) / | h (\mathbf{k}) | \right] $ changes sign under $ \mathbf{k} \to R_{\hat{I}'} \mathbf{k} $ while it is left invariant under $ \mathbf{k} \to R_{\hat{I}} \mathbf{k} $.

If the interacting part
\begin{equation}
 H_\mathrm{int} = \sum_{\sigma_1 \dots \sigma_4} \sum_{\alpha_1 \dots \alpha_4} \int_\mathbb{T} \! d \mathbf{k}_1 \dots d \mathbf{k}_4 \,\,\delta_{\{ \mathbf{k} \}} \,
 f^{\bm{\alpha}} (\mathbf{k}_1,\mathbf{k}_2,\mathbf{k}_3,\mathbf{k}_4) \,\delta_{\sigma_1,\sigma_4} \,\delta_{\sigma_2,\sigma_3}
 \chi^\dagger (\xi_1)\, \chi^\dagger (\xi_2) \, \chi (\xi_3) \, \chi (\xi_4)
\end{equation} 
of the Hamiltonian is expressed in terms of the band pseudo-spinors $ \bm{\chi} $, the coupling function $ f^{\bm{\alpha}} (\mathbf{k}_1,\mathbf{k}_2,\mathbf{k}_3,\mathbf{k}_4) $ is given by
\begin{align} \notag
 f^{\bm{\alpha}} (\mathbf{k}_1,\mathbf{k}_2,\mathbf{k}_3,\mathbf{k}_4) & =
 \frac{U}{2} \left[
 u_{\alpha_1,1} (\mathbf{k}_1) \, u_{\alpha_2,1} (\mathbf{k}_2) \, u_{\alpha_3,1}^\ast (\mathbf{k}_3) \, u_{\alpha_4,1}^\ast (\mathbf{k}_4) \right. \\ \notag
 & \qquad \left. + u_{\alpha_1,2} (\mathbf{k}_1) \, u_{\alpha_2,2} (\mathbf{k}_2) \, u_{\alpha_3,2}^\ast (\mathbf{k}_3) \, u_{\alpha_4,2}^\ast (\mathbf{k}_4) \right] \\ \notag
 & \quad + V \, h (\mathbf{k}_3 - \mathbf{k}_2) \, u_{\alpha_1,1} (\mathbf{k}_1) \, u_{\alpha_2,1} (\mathbf{k}_2) \, u_{\alpha_3,2}^\ast (\mathbf{k}_3) \, u_{\alpha_4,2}^\ast (\mathbf{k}_4) \\
 & \quad + V \, h (\mathbf{k}_3 - \mathbf{k}_2)^\ast \, u_{\alpha_1,2} (\mathbf{k}_1) \, u_{\alpha_2,2} (\mathbf{k}_2) \, u_{\alpha_3,1}^\ast (\mathbf{k}_3) \, u_{\alpha_4,1}^\ast (\mathbf{k}_4) \, .
\end{align} 
 Obviously, the two-particle coupling function behaves trivially under $ \mathbf{k}_i \to R_{\hat{I}} \mathbf{k}_i $, since the transformation matrix $ u (\mathbf{k}) $ does so.
 The behavior under $ \mathbf{k}_i \to R_{\hat{I}'} \mathbf{k}_i $ would also be trivial, if the transformation matrix obeyed $ u (R_{\hat{I}'} \mathbf{k}) = u (\mathbf{k})\, M_{\hat{I}'} $. Unfortunately, this is not
 the case in Eq.~(\ref{eqn:graphene-trans}). However, a trivial behavior of the two-particle coupling function can be enforced by a basis transformation after which we have
\begin{equation}
  u (\mathbf{k}) =  \frac{1}{\sqrt{2}} \left( \begin{array}{ll} -1 & e^{-i \phi (\mathbf{k})}  \\  e^{i \phi (\mathbf{k})} & 1\end{array} \right)
\end{equation} 
for $k_x < 0 $ and
\begin{equation}
  u (\mathbf{k}) = \frac{1}{\sqrt{2}} \left( \begin{array}{ll} e^{i \phi (\mathbf{k})} & -1 \\ 1 & e^{-i \phi (\mathbf{k})} \end{array} \right)
\end{equation} 
elsewhere.  Note that, for the present model, $u (\mathbf{k})$ is continuous everywhere on the BZ in this \emph{natural} band gauge.
\section{Natural Bloch basis for the general case} \label{sec:general}

\subsection{Point-group transformations} \label{sec:trans-rule}

 We now return to the general case, with the notation used in Sec.~\ref{sec:gen-ham}.
In the preceeding examples, the transformation rules for the orbitors in reciprocal space were of the form
\begin{equation}
 \Psi_a (\mathbf{k}) \overset{\hat{O}}{\longrightarrow}
 {\Psi'}_a (R_{\hat{O}} \mathbf{k}) = M_{\hat{O}} (\mathbf{k}) \,
 \Psi_a (\mathbf{k}) \, , \quad \hat{O} \in \mathcal{G}
\end{equation} 
 with $l$-dimensional unitary representation matrices $ M_{\hat{O}} $ of the point group $ \mathcal{G}$. In general, these matrices are wavevector-dependent, as will be explained further below. They must obey the group law
\begin{equation} \label{eqn:group-law}
M_{\hat{C}} (\mathbf{k}) = M_{\hat{B}} (R_{\hat{A}} \mathbf{k}) \, M_{\hat{A}} (\mathbf{k}) \quad  \text{for} \quad  \hat{C} = \hat{B} \hat{A} \, .
\end{equation} 
Let us first discuss the transformation behavior of the quadratic part of the Hamiltonian. Under a point-group operation
$\hat{O}$, the one-particle Hamiltonian density
\begin{equation}
 \Psi^\dagger_a (\mathbf{k}) \, \mathcal{H}_0 (\mathbf{k}) \, \Psi_a (\mathbf{k})
\end{equation} 
 gets mapped to
\begin{equation}
 {\Psi'}^\dagger_a (R_{\hat{O}} \mathbf{k}) \, \mathcal{H}_0 (R_{\hat{O}} \mathbf{k}) \, {\Psi'}_a (R_{\hat{O}} \mathbf{k})
= \Psi^\dagger_a (\mathbf{k}) \, M^\dagger_{\hat{O}} (\mathbf{k}) \, \mathcal{H}_0 (R_{\hat{O}} \mathbf{k})
 \, M_{\hat{O}} (\mathbf{k}) \, \Psi_a (\mathbf{k})  \, .
\end{equation} 
If there now exists a set of representation matrices $M_{\hat{O}} (\mathbf{k})$ with the property
\begin{equation} \label{eqn:invar-quad}
 \mathcal{H}_0 (R_{\hat{O}} \mathbf{k}) = M_{\hat{O}} (\mathbf{k}) \, \mathcal{H}_0 ( \mathbf{k}) \,
M^\dagger_{\hat{O}} (\mathbf{k}) \quad \forall \, \hat{O} \in \mathcal{G} \, ,
\end{equation} 
 the one-particle Hamiltonian density is point-group symmetric. In the presence of interactions, point-group symmetry consequently
 requires the existence of a set of representation matrices that fulfill both Eq.~(\ref{eqn:invar-quad}) and
\begin{align}  \notag
  & \qquad\mathcal{V}_{a_1, \dots, a_{2n}}^{\alpha_1,\dots,\alpha_{2n}}   ( R_{\hat{O}} \mathbf{k}_1, \dots, R_{\hat{O}} \mathbf{k}_n; R_{\hat{O}} \mathbf{k}_{n+1}, \dots, R_{\hat{O}} \mathbf{k}_{2n}) \\  \label{eqn:invar-int}
& =  \sum_{\beta_1, \dots, \beta_{2n}} \left[\prod_{j=1}^{n} \left(M_{\hat{O}} \right)_{\alpha_j,\beta_j}  (\mathbf{k}_j)
\left(M_{\hat{O}}^\dagger \right)_{\beta_{j+n},\alpha_{j+n}}  (\mathbf{k}_{j+n}) \right]
  \mathcal{V}_{a_1, \dots, a_{2n}}^{\beta_1,\dots,\beta_{2n}}   ( \mathbf{k}_1, \dots, \mathbf{k}_n; \mathbf{k}_{n+1}, \dots, \mathbf{k}_{2n}) \, ,
\end{align} 

Clearly, the Hamiltonians for the Emery model and the graphene tight-binding model are point-group invariant as their coupling functions fulfill these relations.
If there are orbitor components
with different spin orientations, one might expect that the (fermionic)
orbitor gets multiplied by $-1 $ under a rotation by $2 \pi $ in analogy to a Dirac spinor.
However, for a charge-conserving
theory, such an additional phase will always cancel and can hence safely be dropped.

 In the above description, point-group symmetry manifests itself in the relations (\ref{eqn:invar-quad}) and (\ref{eqn:invar-int}) for the coupling functions in a second-quantized language.
The precise form of the representation matrices in these equations depends, of course, on the basis.
For example, under a phase transformation
\begin{align}
 \Psi_a^\alpha (\mathbf{k}) &\to e^{i \vartheta_\alpha (\mathbf{k})} \, \Psi_a^\alpha (\mathbf{k})  \, , \\
 \left( {\mathcal{H}_0} (\mathbf{k}) \right)_{\alpha,\beta} & \to e^{i \vartheta_\alpha (\mathbf{k})} \,
 \left( {\mathcal{H}_0} (\mathbf{k}) \right)_{\alpha,\beta} \, e^{-i \vartheta_\beta (\mathbf{k})} \\
 {\mathcal{V}}^{\alpha_1,\dots,\alpha_{2n}}_{a_1,\dots,a_{2n}} (\mathbf{k}_1,\dots,\mathbf{k}_{2n})
 & \to e^{i \vartheta_{\alpha_1} (\mathbf{k})} \dots e^{i \vartheta_{\alpha_n} (\mathbf{k})} \,
 {\mathcal{V}}^{\alpha_1,\dots,\alpha_{2n}}_{a_1,\dots,a_{2n}} (\mathbf{k}_1,\dots,\mathbf{k}_{2n})
 \, e^{-i \vartheta_{\alpha_{n+1}} (\mathbf{k})} \dots e^{-i \vartheta_{\alpha_{2n}} (\mathbf{k})}
\end{align} 
in the orbital language,
momentum-independent representation matrices $ M_{\hat{O}} $ may be rendered momentum-dependent according to
\begin{equation} \label{eqn:M-phas}
 \left( {M_{\hat{O}}} (\mathbf{k}) \right)_{\alpha,\beta} \to
 e^{i \vartheta_\alpha (R_{\hat{O}} \mathbf{k})} \, \left( M_{\hat{O}} (\mathbf{k}) \right)_{\alpha,\beta} \, e^{-i \vartheta_\beta (\mathbf{k})} \, .
\end{equation} 
 As the reader may easily verify, this transformation does not affect the group law in Eq.~(\ref{eqn:group-law}). A multiband model of the type given in Eq.~(\ref{eqn:gen-model}) is therefore
 invariant under point-group operations irrespective of the choice of the phases $ \vartheta_\alpha (\mathbf{k}) $.

One may therefore wonder whether Eqs.~(\ref{eqn:invar-quad}) and (\ref{eqn:invar-int}) can be derived by postulating vanishing commutators as a starting point.
 This can be accomplished as follows.
 For a given $\mathcal{G}$-symmetric  Hamiltonian,
 there exists a set of unitary operators $ D_{\hat{O}} $ which is isomorphic to the point group $\mathcal{G}$.
 They act on an arbitrary one-particle state $ | \psi \rangle $ with wave function $ \langle \mathbf{r} | \psi \rangle $ in position representation according to
 \begin{equation}
  \left\langle \left. \mathbf{r} \left| D_{\hat{O}} \right| \psi \right. \right\rangle = \left\langle \left. R_{\hat{O}}^{-1} \mathbf{r} \right| \psi \right\rangle \, .
 \end{equation} 
 If we require the Bloch states to transform as
\begin{equation} \label{eqn:trans-bloch}
 D_{\hat{O}} \left| \phi_a^\alpha (\mathbf{k}) \right\rangle = \sum_\beta \left(M^\dagger_{\hat{O}} \right)_{\alpha,\beta} (\mathbf{k})
\left| \phi_a^\beta (R_{\hat{O}}\mathbf{k}) \right\rangle \, ,
\end{equation} 
the above representation matrices $ M_{\hat{O}} (\mathbf{k}) $ are given by
\begin{equation}
 \left( M_{\hat{O}} (\mathbf{k}) \right)_{\alpha,\beta} = \int \! d \mathbf{r} \, \left\langle R_{\hat{O}} \mathbf{r} \left| \phi_a^\alpha (R_{\hat{O}} \mathbf{k}) \right. \right\rangle
\left\langle \left. \phi_a^\beta (\mathbf{k}) \right| \mathbf{r} \right\rangle
\end{equation} 
 for arbitrary $a$. Under a phase transformation of the hybridizing Bloch basis, Eq.~(\ref{eqn:M-phas}) is recovered from this formula.
 For a point-group symmetric model, the representation operators $D_{\hat{O}} $ commute with the Hamiltonian
\begin{equation} \label{eqn:commu}
 \left[ D_{\hat{O}} , H \right] = 0 \quad \forall \, \hat{O} \in \mathcal{G}
\end{equation} 
 and with its coupling functions which are complex numbers.
 Consequently, the point-group symmetries must be encoded in the behavior of the field operators under
\begin{equation}
  \Psi_a (\mathbf{k}) \to D_{\hat{O}} \, \Psi_a (\mathbf{k})\, D_{\hat{O}}^\dagger \, .
\end{equation} 
 From  Eq.~(\ref{eqn:trans-bloch}), it follows that
\begin{equation}
  D_{\hat{O}}\, \Psi_a (\mathbf{k})\, D_{\hat{O}}^\dagger   = M^\dagger_{\hat{O}} (\mathbf{k}) \, \Psi_a (R_{\hat{O}} \mathbf{k}) \, ,
\end{equation} 
 since the vacuum reference state $ | 0 \rangle = D_{\hat{O}} | 0 \rangle $ is mapped onto itself under all point-group
 operations.
 Together with Eq.~(\ref{eqn:commu}), this implies the validity of Eqs.~(\ref{eqn:invar-quad}) and (\ref{eqn:invar-int}).
\subsection{Basis transformations in the band language} \label{sec:gen-band}

 We are now in a position to look at our general model in the band language, i.e.\ we switch from orbitors $ \Psi_a (\mathbf{k}) $ to band pseudo-spinors
 $ \bm{\chi}_a (\mathbf{k}) = u (\mathbf{k}) \, \Psi_a (\mathbf{k}) $.
The orbital-to-band transformation $u (\mathbf{k}) $ is chosen such that
it renders the one-particle coupling-function
\begin{equation}
  \mathcal{B} (\mathbf{k}) = u(\mathbf{k}) \, \mathcal{H}_0 (\mathbf{k}) \, u^\dagger (\mathbf{k})
\end{equation} 
diagonal.
Since the eigenvalues of $ \mathcal{H}_0 $ will be invariant under a unitary transformation and since
we have such a transformation on the right-hand side of Eq.~(\ref{eqn:invar-quad}), the band labels can be chosen such that
\begin{equation} \label{eqn:band-inv}
 \mathcal{B}(\mathbf{k}) = \mathcal{B}( R_{\hat{O}} \mathbf{k})
\end{equation} 
 holds. So the point-group symmetry of the Hamiltonian already implies that the band dispersion transforms trivially under $ \mathbf{k} \to R_{\hat{O}} \mathbf{k} $.
 In the following, the bands will always be labeled in a way that guarantees Eq.~(\ref{eqn:band-inv}).

Let us ignore the interactions for a moment.
Then different (unitary) representation matrices $ \tilde{M}_{\hat{O}} (\mathbf{k}) $ could have been chosen in Eq.~(\ref{eqn:invar-quad}),
if
 \begin{equation} \label{eqn:A}
  \tilde{M}_{\hat{O}} (\mathbf{k})  = M_{\hat{O}} (\mathbf{k}) \, A_{\hat{O}} (\mathbf{k})
 \end{equation} 
with a unitary matrix $ A_{\hat{O}} (\mathbf{k}) $  commuting with $ \mathcal{H}_0 (R_{\hat{O}} \mathbf{k}) $.
Note that replacing $ M_{\hat{O}} (\mathbf{k}) $ by $ \tilde{M}_{\hat{O}} (\mathbf{k}) $ may violate the transformation rule (\ref{eqn:invar-int}) for $ H_\mathrm{int} $.
We will now show that the two $l$-dimensional representations of $\mathcal{G} $ with representation matrices
$ M_{\hat{O}} (\mathbf{k}) $ and $ \tilde{M}_{\hat{O}} (\mathbf{k}) $, respectively,
 are connected by a basis transformation between basis sets of non-hybridizing Bloch states. Such a transformation corresponds to a different choice of the eigenvectors of $ \mathcal{H}_0 (\mathbf{k}) $, i.e.\ the orbital-to-band
matrix $ u_{\alpha \beta}  (\mathbf{k}) $ is substituted by
 $ \tilde{u}_{\alpha \beta} (\mathbf{k}) = e^{-i \varphi_\alpha (\mathbf{k})} \, u_{\alpha \beta}  (\mathbf{k}) $.
 Due to momentum conservation, the one-particle coupling function $ \mathcal{B} $ in the band language remains
unaffected by such transformations. This may be interpreted as an \emph{emergent} local U(1) gauge symmetry at
a Fermi liquid fixed point.  \cite{sun-fradkin}
 We emphasize that, away from such a fixed-point, this gauge symmetry is violated or that, in other words, the coupling functions of a non-vanishing interaction term may change under a basis transformation.

Clearly, the point-group symmetry Eq.~(\ref{eqn:band-inv}) of the band dispersion implies
\begin{equation}
 \mathcal{H}_0 (\mathbf{k}) = u^\dagger(\mathbf{k}) \, u (R_{\hat{O}} \mathbf{k}) \,\, \mathcal{H}_0 (R_{\hat{O}} \mathbf{k}) \, \,u^\dagger (R_{\hat{O}} \mathbf{k}) \, u (\mathbf{k}) \, .
\end{equation} 
and therefore the representation matrices
\begin{equation} \label{eqn:rep-quad}
  M_{\hat{O}} (\mathbf{k}) =  u^\dagger (R_{\hat{O}} \mathbf{k}) \,  u (\mathbf{k})
\end{equation} 
satisfy Eq.~(\ref{eqn:invar-quad}).
On the other hand, changing the phase of the bands gives rise to representation matrices
\begin{equation} \label{eqn:isomorph}
  \tilde{M}_{\hat{O}} (\mathbf{k})  = u^\dagger(R_{\hat{O}} \mathbf{k}) \, P_{\hat{O}}  (\mathbf{k}) \, u (\mathbf{k}) \, ,
\end{equation} 
where $ \left( P_{\hat{O}}\right)_{\alpha \beta} (\mathbf{k}) = \delta_{\alpha \beta} \, e^{i \left[ \varphi_\beta (R_{\hat{O}} \mathbf{k}) -
 \varphi_\beta (\mathbf{k}) \right]} $. Being a product of unitary matrices, the $\tilde{M}_{\hat{O}} (\mathbf{k}) $ are themselves unitary. It is now straightforward to show that the mapping given by Eq.~(\ref{eqn:isomorph})
 is an isomorphism between two $l$-dimensional representations of $ \mathcal{G} $.
 Namely,  the group law (\ref{eqn:group-law}) also holds for the new representation matrices in Eq.~(\ref{eqn:isomorph}) for arbitrary phases $ \varphi_\beta (\mathbf{k}) $, since $ P_{\hat{B}} (\mathbf{k}) \, P_{\hat{A}} (R_{\hat{B}} \mathbf{k}) = P_{\hat{C}} (\mathbf{k}) $ for $ \hat{C} = \hat{B} \hat{A} $. Furthermore, one finds that Eq.~(\ref{eqn:A})
 is fulfilled for
\begin{equation}
  A_{\hat{O}} (\mathbf{k})  = u^\dagger(\mathbf{k}) \, P_{\hat{O}} (\mathbf{k}) \, u (\mathbf{k}) \, .
\end{equation} 
As far as the one-particle Hamiltonian is concerned, a basis transformation $ u (\mathbf{k}) \to \tilde{u} (\mathbf{k}) $ just maps a representation of $\mathcal{G} $ with matrices $M_{\hat{O}} (\mathbf{k}) = u^\dagger (R_{\hat{O}} \mathbf{k}) \, u (\mathbf{k}) $ onto one with matrices
 $\tilde{M}_{\hat{O}} (\mathbf{k}) =\tilde{u}^\dagger (R_{\hat{O}} \mathbf{k}) \, \tilde{u} (\mathbf{k})  $. This implies that, for any choice of the $\varphi_\beta (\mathbf{k}) $, there exists an $l$-dimensional representation of $ \mathcal{G} $, with which the orbital-to-band matrix transforms under a point-group operation, i.e.
\begin{equation}
 \tilde{u} (\mathbf{k}) = \tilde{u} (R_{\hat{O}} \mathbf{k}) \, \tilde{M}_{\hat{O}} (\mathbf{k}) \, , \quad \forall \, \hat{O} \in \mathcal{G} \, .
\end{equation} 
 The one-particle coupling function $ \mathcal{H}_0 (\mathbf{k}) $ in the orbital language may be transformed with each of these representations under $ \mathbf{k} \to
R_{\hat{O}} \mathbf{k} $.

Now we are in a position to address the question of a sensible fixing of the phases of the bands in the presence of interactions.
Changing these phases then replaces $ M_{\hat{O}} (\mathbf{k}) $ by $ \tilde{M}_{\hat{O}} (\mathbf{k}) $ in the transformation rule (\ref{eqn:invar-quad}) for the one-particle Hamiltonian, but the transformation rule (\ref{eqn:invar-int}) for the interactions may not hold with the new representation matrices $ \tilde{M}_{\hat{O}} (\mathbf{k}) $ in general.
A particular basis of non-hybridizing Bloch states corresponding to $ u (\mathbf{k}) $ shall henceforth be called \emph{natural} if also $ H_\mathrm{int} $ remains invariant under
\begin{equation}
 \Psi_a (\mathbf{k}) \to {\Psi'}_a (R_{\hat{O}} \mathbf{k}) = u^\dagger (R_{\hat{O}} \mathbf{k}) \, u (\mathbf{k}) \, \Psi_a (\mathbf{k})
\end{equation} 
for all operations $ \hat{O} \in \mathcal{G} $. This means that, if one finds a given Hamiltonian to transform according to  Eqs.~(\ref{eqn:invar-quad}) and (\ref{eqn:invar-int})
with  representation matrices $ M_{\hat{O}} (\mathbf{k}) $, in natural Bloch basis
\begin{equation} \label{eqn:gauge-fix}
u (R_{\hat{O}} \mathbf{k}) = u (\mathbf{k}) \,  M^\dagger_{\hat{O}} (\mathbf{k})
\end{equation} 
must be fulfilled. Formally, this equation is equivalent to Eq.~(\ref{eqn:rep-quad}). However, a condition on $u(\mathbf{k})$ is imposed for given representation matrices
 $ M_{\hat{O}} (\mathbf{k}) $ here.

That such a natural basis must always exist, can be seen as follows. When $ \mathcal{H}_0 (R_{\hat{O}} \mathbf{k}) $ in Eq.~(\ref{eqn:invar-quad}) is diagonalized, the symmetry of the band dispersion (\ref{eqn:band-inv}) implies that $ \mathcal{H}_0 (\mathbf{k}) $ is as well diagonalized by $ u (R_{\hat{O}} \mathbf{k}) \, M_{\hat{O}} (\mathbf{k}) $,
 which therefore can be identified as a possible choice of $ u(\mathbf{k}) $ for given $ u (R_{\hat{O}} \mathbf{k}) $.
If $u(\mathbf{k}) $ is given at some point $ \mathbf{k} = \mathbf{q} $, the group law (\ref{eqn:group-law}) then ensures that, within a natural basis, $ u(\mathbf{k}) $ is uniquely
defined on the star of $\mathbf{q}$, i.e.\ at $ \mathbf{k} = R_{\hat{O}} \mathbf{q} $ $ \forall \hat{O} \in \mathcal{G}$.
 For a particular model, however, there
are infinitely many natural bases corresponding to different $l$-dimensional representations that all satisfy Eqs.~(\ref{eqn:invar-quad}) and
(\ref{eqn:invar-int}). Namely, if we start from a natural basis and perform a phase transformation
\begin{equation}
 u_{\alpha,\beta} (\mathbf{k}) \to u_{\alpha,\beta} (\mathbf{k}) \, e^{- i \varphi_\beta(\mathbf{k}) } \, , \quad
\text{where} \quad \varphi_\beta (R_{\hat{O}} \mathbf{k}) = \varphi_\beta(\mathbf{k}) \quad \forall \, \hat{O} \in \mathcal{G} \, ,
\end{equation} 
the properties of a natural basis are preserved.
\subsection{Consequences of the existence of a natural basis} \label{sec:gen-natural}

 For the above examples of the Emery model and for graphene, we have seen that, expressed in a natural basis, the coupling functions of the interaction transform trivially under all point-group operations.
 In this place, the reader may probably ask whether this
also holds for the general Hamiltonian discussed in this section. Let us therefore look at the transformation properties of the coupling functions $ \mathcal{F} $ of the interaction in Eq.~(\ref{eqn:band-inter}), substitute $ \mathbf{k}_i $ by $ R_{\hat{O}} \mathbf{k}_i $ and insert Eqs.~(\ref{eqn:invar-int}) and (\ref{eqn:gauge-fix}). The representation matrices then cancel and
 therefore the $n$-particle coupling function indeed transforms trivially, i.e.
\begin{equation}
  \mathcal{F}_{a_1, \dots, a_{2n}}^{\alpha_1,\dots,\alpha_{2n}}  ( R_{\hat{O}} \mathbf{k}_1, \dots, R_{\hat{O}} \mathbf{k}_n; R_{\hat{O}} \mathbf{k}_{n+1}, \dots, R_{\hat{O}} \mathbf{k}_{2n})
  = \mathcal{F}_{a_1, \dots, a_{2n}}^{\alpha_1,\dots,\alpha_{2n}}  ( \mathbf{k}_1, \dots, \mathbf{k}_n; \mathbf{k}_{n+1}, \dots, \mathbf{k}_{2n})  \, .
\end{equation} 
 Consequently, the full Hamiltonian is invariant under
\begin{equation} \label{eqn:trans-nat}
 \bm{\chi}_a (\mathbf{k}) \to \bm{\chi'}_a (R_{\hat{O}} \mathbf{k}) = \bm{\chi}_a (\mathbf{k})  \, .
\end{equation} 
 This corresponds to the trivial point-group behavior of the natural basis states
\begin{equation} \label{eqn:pg-basis}
 D_{\hat{O}} \left| \chi_a^\alpha (\mathbf{k})\right\rangle =\left| \chi_a^\alpha (R_{\hat{O}} \mathbf{k})\right\rangle \, ,
\end{equation} 
 which follows from Eqs.~(\ref{eqn:trans-bloch}) and (\ref{eqn:gauge-fix}).

But also in cases where a natural basis seems a suboptimal choice, the existence of a natural basis has consequences that may simplify analytical and/or numerical calculations. Let us therefore start from a natural basis with states $ \left| \chi_a^\alpha ( \mathbf{k})\right\rangle $ and switch to a non-natural one with states
\begin{equation} \label{eqn:non-awkward}
 \left| \eta_a^\alpha( \mathbf{k}) \right\rangle = e^{i \varphi_\alpha ( \mathbf{k}) } \left| \chi_a^\alpha ( \mathbf{k})\right\rangle = {\bm{\eta}^\alpha_a}^\dagger (\mathbf{k})
 \left| 0 \right\rangle
\end{equation} 
 and pseudo-spinor fields  $ \bm{\eta}_a (\mathbf{k}) $.
 If there are degeneracies in the bands at some points, also non-hybridizing Bloch states exist that violate Eq.~(\ref{eqn:non-awkward}).
 Namely, at these points, the degenerate bands may get mixed. At all other points, Eq.~(\ref{eqn:non-awkward}) is of course still respected.
 We will refer to such a band basis as an \emph{awkward} one. This name seems already justified since band degeneracies typically occur at singular points,
 and since the transformation from
 natural to awkward Bloch states would hence be discontinuous at the band degeneracies. (Note, however, that this does not imply the continuity of the orbital-to-band
 transformation for non-awkward states on the whole BZ.) In addition, awkward Bloch states may have other pathological properties, as will become clear in the following.

 But let us first look at the point-group properties in non-awkward non-natural bases.
Transforming Eqs.~(\ref{eqn:trans-nat}) and (\ref{eqn:pg-basis}) according to Eq.~(\ref{eqn:non-awkward}) yields
\begin{equation}
 D_{\hat{O}} \left| \eta_a^\alpha (\mathbf{k})\right\rangle = e^{i \left[\varphi_\alpha (\mathbf{k}) -\varphi_\alpha (R_{\hat{O}} \mathbf{k}) \right] } \,
 \left| \eta_a^\alpha (R_{\hat{O}} \mathbf{k})\right\rangle \, ,
 \end{equation} 
and
\begin{equation}
 \bm{\eta}_a (\mathbf{k}) \to {\bm{\eta'}_a} (R_{\hat{O}} \mathbf{k}) = N_{\hat{O}} (\mathbf{k}) \, \bm{\eta}_a (\mathbf{k})\, , \qquad
\left( N_{\hat{O}} (\mathbf{k}) \right)_{\alpha \beta} = \delta_{\alpha \beta} \, e^{i \left[\varphi_\alpha (R_{\hat{O}} \mathbf{k}) -\varphi_\alpha (\mathbf{k}) \right] } \,
\end{equation} 
 respectively.
 So non-awkward non-hybridizing Bloch states  with arbitrarily chosen $\varphi_\alpha$ in Eq.~(\ref{eqn:non-awkward}) acquire a phase factor under a point-group operation.
 Hence, the representation matrices $M_{\hat{O}} (\mathbf{k}) $ in the orbital language can be said to have \emph{diagonal} counterparts $N_{\hat{O}} (\mathbf{k}) $
 in the band language with
a non-awkward basis or, in other words, non-awkward bands transform with one-dimensional representations of the point group. At points, where the bands are non-degenerate,
 the latter statement is already well known (cf.\ Chapter~8-4 of Ref.~\onlinecite{tinkham-groupth}) without the notion of a natural basis.
 For a further discussion and an alternative proof of the existence of a natural gauge in the absence of degeneracies away from points of high symmetry, the reader
 shall be referred to Appendix~\ref{sec:alt-proof-nat}.

 In a similar way as for the fields, the orbital-to-band
 transformation matrix $ v $ corresponding to the non-awkward states of Eq.~(\ref{eqn:non-awkward})  transforms with a phase factor according to
\begin{equation}
v (R_{\hat{O}} \mathbf{k}) = N_{\hat{O}} (\mathbf{k}) \, v (\mathbf{k}) \,  M^\dagger_{\hat{O}} (\mathbf{k}) \, ,
\end{equation} 
 which follows straightforwardly from Eq.~(\ref{eqn:gauge-fix}). Together with Eqs.~(\ref{eqn:band-inter}) and (\ref{eqn:invar-int}), this implies that the $n$-particle coupling function enjoys a rather simple point-group behavior, namely
\begin{align}  \notag
  & \qquad\mathcal{F}_{a_1, \dots, a_{2n}}^{\alpha_1,\dots,\alpha_{2n}}   ( R_{\hat{O}} \mathbf{k}_1, \dots, R_{\hat{O}} \mathbf{k}_n; R_{\hat{O}} \mathbf{k}_{n+1}, \dots, R_{\hat{O}} \mathbf{k}_{2n}) \\  \label{eqn:int-nawk}
& =  \exp \left\{ i \sum_{j=1}^{2n}  s_j \left[\varphi_{\alpha_j} (R_{\hat{O}} \mathbf{k}_j) -\varphi_{\alpha_j} (\mathbf{k}_j) \right] \right\} \,
  \mathcal{F}_{a_1, \dots, a_{2n}}^{\alpha_1,\dots,\alpha_{2n}}   ( \mathbf{k}_1, \dots, \mathbf{k}_n; \mathbf{k}_{n+1}, \dots, \mathbf{k}_{2n}) \\
& =  \sum_{\beta_1, \dots, \beta_{2n}} \left[\prod_{j=1}^{n} \left(N_{\hat{O}} \right)_{\alpha_j,\beta_j}  (\mathbf{k}_j)
\left(N_{\hat{O}}^\dagger \right)_{\beta_{j+n},\alpha_{j+n}}  (\mathbf{k}_{j+n}) \right]
  \mathcal{F}_{a_1, \dots, a_{2n}}^{\beta_1,\dots,\beta_{2n}}   ( \mathbf{k}_1, \dots, \mathbf{k}_n; \mathbf{k}_{n+1}, \dots, \mathbf{k}_{2n}) \, ,
\end{align} 
 where $ s_j = +1$ for $ j \leq n $ and $ s_j = -1 $ for $ j > n $.
 The non-natural basis sets for the Emery model and graphene given in Sec.~\ref{sec:Emery-natural} and \ref{sec:graphene-natural}, respectively, are non-awkward and hence
 enjoy these properties.

If one finds a way how to deal with the phases in Eq.~(\ref{eqn:int-nawk}), it may also be convenient to work in a non-awkward, non-natural basis.
 This approach was pursued in fRG studies \cite{wang-2009,wang-2010,Thomale-Platt-pnictides,platt-first,pnictides-gap,sidplatt,kiesel2,sruo}
 of multiband models with Fermi surface patching.
 In awkward bases, Eq.~(\ref{eqn:int-nawk}) would be violated at points with band degeneracies \emph{away} from the origin (or at other points of high symmetry), justifying the name we chose for those bases.
 In practice, working in a non-natural band basis may hence require a careful treatment of such band degeneracies.
 If there is only a degeneracy at the origin, the transformation rule in Eq.~(\ref{eqn:int-nawk}) still holds, since it is trivially fulfilled at this point.
\subsection{Construction of a natural Bloch basis}

 We now comment on more practical aspects of a natural fixing of the phases of the bands for a given model. In doing so, one has to fix  $l$ phases, one for each band. In contrast, the condition~(\ref{eqn:gauge-fix}) for a
 natural basis corresponds to $l^2$ constraints for $l$ variables. Hence, there must be some redundancy, as a natural basis must exist as shown above.
 (The reason for this redundancy lies in the diagonalizability of $ u (\mathbf{k}) $, which therefore has only $l$ \emph{independent} entries.)

 If no zeros appear in $u (\mathbf{k}) $, the phases of the band can be fixed to a natural basis by taking an arbitrary row in Eq.~(\ref{eqn:gauge-fix}), i.e.\ by choosing an arbitrary orbital index.
 For the example of graphene, the orbital-to-band matrix has no zeros and hence one may proceed in this way.
 In the case of the Emery model, however, the flat band has no $d$-orbital component and hence the first row of Eq.~(\ref{eqn:gauge-fix}) only fixes the phases of the other bands.
 If one allows for oxygen-oxygen hopping, which was absent in Sec.~\ref{sec:Emery}, the situation is different. Then there is no pure $p$-band any more, and  all entries in
 $ u_{\alpha 1} (\mathbf{k}) $ are non-zero except at the origin, were Eq.~(\ref{eqn:gauge-fix}) is trivial. The first line
$ u_{\beta 1} (R_{\hat{O}} \mathbf{k}) = u_{\beta 1} (\mathbf{k}) $ can then be fulfilled
 by requiring
\begin{equation}
 u_{\beta 1} (\mathbf{k}) \,\, \in \mathbb{R}  , \, > 0 \quad \forall \beta,\mathbf{k} \, ,
\end{equation} 
as the $u$ can be chosen real.

 For the general case, the following construction scheme seems appropriate.
\begin{enumerate}[i)]
 \item Select a minimal sector $\mathbb{S} $ of the BZ $\mathbb{B}$, i.e.\ the smallest set of points from which $\mathbb{B}$ can be generated according to
\begin{equation}
 \mathbb{B} = \left\{ \mathbf{p} \left| \mathbf{p} = R_{\hat{O}} \mathbf{q} , \mathbf{q} \in \mathbb{S}, \hat{O} \in \mathcal{G} \right. \right\} \, .
\end{equation} 
 For $C_{4v}$, for example, $ \mathbb{S}$ corresponds to $1/8$ of the BZ.
 \item Choose the band phases in the orbital-to-band transformation $u (\mathbf{k}) $ arbitrarily on $\mathbb{S}$.
 \item Determine $u (R_{\hat{O}} \mathbf{k}) $ from $ u (\mathbf{k}) $ for $ \mathbf{k} \in \mathbb{S} $ via Eq.~(\ref{eqn:gauge-fix}).
 Since $ M_{\hat{O}} (\mathbf{k}) = 1 $ for $ \hat{O} \mathbf{k} = \mathbf{k} $, this scheme is free of inconsistencies.
\end{enumerate}
 
 Before we conclude, let us briefly comment on the locality of the basis states, which plays an important role in dynamical mean-field theory and its extensions.
Under a transformation from a non-natural to a natural Bloch basis,
there seems to be no generic tendency in the localization properties of the corresponding (non-hybridizing) Wannier states.
These latter states are simply obtained from the respective Bloch basis by a Fourier transformation. 
A fast decay of the hopping parameters with increasing distance may be regarded as a hallmark of their locality.
Since the band dispersion is invariant under this transformation, the coefficients of its hopping-parameter expansion remain unaffected, and the localization
properties seem not to change much in this picture.

\section{Summary and outlook} \label{sec:concl}

 In this work, we have discussed how point-group symmetries are represented in a large class of  multiband models -- as well in the orbital as in the band language.
 We have presented the Emery model without oxygen-oxygen hopping and a graphene tight-binding model as examples with coupling functions that are are still analytically accessible in the band language.
 In the orbital language, not only the momentum, but also the orbital quantum numbers transform with representation matrices of the point group. Under a transformation from one basis of hybridizing Bloch functions to another, these representation matrices are
mapped onto equivalent ones with a different momentum dependence.

 If one now switches to the band language, the band dispersion transforms trivially under all point-group operations. On a one-particle level, this gives rise to an invariance under a
 transformation to another basis set of non-hybridizing Bloch states,
reflecting the arbitrariness of the phases of eigenvectors.
 Such a transformation corresponds to an isomorphism between representations matrices in the orbital language. The interactions, however, may not transform in the same way with all of these representations. In the band language, this implies that
the interaction can be more conveniently expressed in some bases. We have shown that the vertex functions of the interactions simply transform by a rotation of their
 momentum arguments for such a \emph{natural basis}, without additional phases or a reordering of the bands.

 The fixing of the band phases for the natural basis may render the vertex functions discontinuous at momenta on symmetry elements such as inversion centers, mirror axes or planes.
 However, if one finds a way how to deal with these discontinuities, the point-group symmetries can now be exploited in the numerical calculation of Feynman diagrams.
 Moreover, the existence of natural bases gives rise to a simple point-group behavior if a non-natural basis is used. 
 With the exception of awkward basis sets,
the vertex functions then acquire momentum-dependent phases attached to their external legs under a point-group transformation of their momentum arguments.
 Instead of working in a natural basis,
 one may therefore also exploit the point-group symmetry by keeping track of these phases.

 Either way, the computational effort of fRG studies may be lowered considerably, in particular for  two-dimensional systems with a sixfold symmetry such multilayer graphene \cite{BLG-frg,BLG-AF,TLG-frg} and electrons on a kagome lattice.  \cite{kiesel2,ch-dec-kagome}
 Currently, two of us are preparing a publication on an fRG study of the Emery model with \emph{non-zero} oxygen-oxygen hopping and at weak coupling.
 We also hope that the findings of this work may also be helpful in the context of other analytical or numerical many-particle methods based on a reciprocal
 space description. This should be possible for perturbative and self-consistent methods, where vertex functions are an important building block.
%

%%%%%%%%%%%%%%%%%%%%%%%%%%%%%%%%%%%%%%%%%%

\acknowledgements{Acknowledgments}

 We thank C.~Platt, D.~Sanchez~de~la~Pe\~{n}a, M.~M.~Scherer, and M.~Schmidt for fruitful discussions. S.~M.\ was funded by the German Research Foundation DFG via FOR 723.
 Q.-H.~W.\ acknowledges support by NSFC (under grant No.~11023002) and the Ministry of Science and Technology of China (under grant No.~2011CBA00108 and 2011CB922101).
\begin{appendix}
 \section{Alternative proof of the existence of a natural basis} \label{sec:alt-proof-nat}

 \subsection{Non-degenerate bands}

 In this Appendix, we present a short, alternative proof the existence of a natural basis. Let us first only allow for band degeneracies at high-symmetry points,
where the little group equals the full point group. After considering this more special case, let us generalize this
 proof to band structures with degeneracies at arbitrary points.

 It is textbook knowledge (see, for example, Chapter 8-4 of Ref.~\onlinecite{tinkham-groupth}) that non-degenerate bands transform with one-dimensional irreducible representations
 of the point group. On a more formal level, this means that band pseudo-spinors $ \bm{\eta} $ behave according to
\begin{equation} \label{eqn:non-degenerate}
 \bm{\eta}_a (\mathbf{k}) \to {\bm{\eta'}_a} (R_{\hat{O}} \mathbf{k}) = N_{\hat{O}} (\mathbf{k}) \, \bm{\eta}_a (\mathbf{k})\, , \qquad
\left( N_{\hat{O}} (\mathbf{k}) \right)_{\alpha \beta} = \delta_{\alpha \beta} \, e^{i \theta_{\hat{O}}^\alpha (\mathbf{k}) } \, .
\end{equation} 
The phase shifts $ \theta_{\hat{O}}^\alpha (\mathbf{k}) $ obey the group law
\begin{equation} \label{eqn:glaw-phases}
 \theta_{\hat{C}}^\alpha (\mathbf{k}) = \theta_{\hat{B}}^\alpha (R_{\hat{A}} \mathbf{k}) + \theta_{\hat{A}}^\alpha (\mathbf{k}) \quad \text{for} \quad \hat{C} = \hat{B} \hat{A}  \, .
\end{equation} 
 The transformation rule~(\ref{eqn:non-degenerate}) is also fulfilled in the presence of band degeneracies at high-symmetry points $\mathbf{q}$,
 where $ \mathbf{q} = R_{\hat{O}} \mathbf{q} $
 $ \forall \hat{O} \in\mathcal{G} $. At these points, the phase shifts $  \theta_{\hat{O}}^\alpha (\mathbf{k}) $ must vanish for all bands $\alpha $ and all point-group
 operations $\hat{O} \in\mathcal{G} $.
 If Eq.~(\ref{eqn:non-degenerate}) holds everywhere on the BZ, one may always introduce phases $ \varphi_\alpha (\mathbf{k}) $ such that
 $ \theta_{\hat{O}}^\alpha (\mathbf{k}) = \varphi_\alpha (R_{\hat{O}} \mathbf{k}) - \varphi_\alpha (\mathbf{k}) $.
 After $ \varphi_\alpha (\mathbf{k}) $ is fixed at some arbitrary point $ \mathbf{k} $, it is uniquely
 defined on the star of $\mathbf{k}$ by virtue of the group law~(\ref{eqn:glaw-phases}).

 The consequences of this are twofold. For one thing, the vertex functions now transform with phases factors attached to their external legs, i.e.\ one has
$ \mathcal{B} (R_{\hat{O}} \mathbf{k}) = \mathcal{B} (\mathbf{k}) $ and Eq.~(\ref{eqn:int-nawk}) holds for the interaction.
 Moreover, one may now perform a phase transformation
\begin{equation}
 \bm{\eta}_a^\alpha (\mathbf{k}) \to\bm{\chi}_a^\alpha (\mathbf{k}) = e^{- i \varphi_\alpha (\mathbf{k})} \, \bm{\eta}_a^\alpha (\mathbf{k}) \, ,
\end{equation} 
 which renders all these phase equal to unity. The corresponding Bloch basis is then a natural one, since it transforms according to Eq.~(\ref{eqn:trans-nat}) under
 point-group operations.
 \subsection{General case}

 We now allow for the bands to touch in arbitrary locations of the BZ, provided that the point-group symmetry of the band structure is still respected.
 This means, that if there is a degeneracy at $ \mathbf{k} $, this degeneracy can also be found on the star of $\mathbf{k} $.
 Eq.~(\ref{eqn:non-degenerate}) now generalizes to
\begin{equation}
 \bm{\eta}_a (\mathbf{k}) \to {\bm{\eta'}_a} (R_{\hat{O}} \mathbf{k}) = \tilde{N}_{\hat{O}} (\mathbf{k}) \, \bm{\eta}_a (\mathbf{k})\, ,
\end{equation} 
 with unitary representation matrices $\tilde{N}_{\hat{O}} (\mathbf{k}) $ which are equal to unity at high-symmetry points and which decay into irreducible blocks.
If the bands do not touch at $\mathbf{k}$, these blocks are one-dimensional, just as before.
 At band degeneracies, however, they may contain higher-dimensional irreducible blocks mixing the degenerate bands under point-group operations.
 The group law for the new representation matrices
\begin{equation}
 \tilde{N}_{\hat{C}} (\mathbf{k}) = \tilde{N}_{\hat{B}} (R_{\hat{A}} \mathbf{k}) \,  \tilde{N}_{\hat{A}} (\mathbf{k}) \quad \text{for} \quad \hat{C} = \hat{B} \hat{A}  \, ,
\end{equation} 
 allows for a decomposition
\begin{equation} \label{eqn:decomp}
  \tilde{N}_{\hat{O}} (\mathbf{k}) =w^\dagger ( R_{\hat{O}}\mathbf{k}) \, w (\mathbf{k})
\end{equation} 
 with unitary matrices $w $. This decomposition is not unique and Eq.~(\ref{eqn:decomp}) rather imposes a constraint on possible choices of $w(\mathbf{k})$.
 If one assumes some arbitrary unitary $w$ at $\mathbf{k} $, $w$ is uniquely defined on the star of $\mathbf{k}$ according to this constraint.
 Moreover, if $w$ has the block structure of the $\tilde{N}_{\hat{O}} $ at $\mathbf{k}$, this also holds on the star of $\mathbf{k}$.
 One may now perform a basis transformation
\begin{equation}
 \bm{\eta}_a^\alpha (\mathbf{k}) \to\bm{\chi}_a^\alpha (\mathbf{k}) = w (\mathbf{k}) \, \bm{\eta}_a^\alpha (\mathbf{k})
\end{equation} 
 with $w$ decomposing the representation matrices $\tilde{N}_{\hat{O}} $.
 Such a basis transformation then renders the $\tilde{N}$ all equal to a unit matrix and Eq.~(\ref{eqn:trans-nat}) holds.
 If one now further requires that the $\tilde{N}_{\hat{O}} (\mathbf{k}) $ and $w (\mathbf{k}) $
 have the same irreducible block structure everywhere on the BZ, only degenerate bands get mixed.
 Hence, the one-particle Hamiltonian is still diagonal after the transformation and the new basis is a natural one.
\end{appendix}
%%%%%%%%%%%%%%%%%%%%%%%%%%%%%%%%%%%%%%%%%%

%=================================================================
% References:  Variant B
%=================================================================
% Use the following option to include external BibTeX files:
\bibliography{biblio}
\end{document}